\documentclass[draftclsnofoot, letterpaper, onecolumn, 12pt]{IEEEtran}
\usepackage{multirow}
\usepackage{booktabs}
\usepackage{amsfonts}
\usepackage{amsmath,cases}
\usepackage{amssymb}
\usepackage{graphicx}
\usepackage{cite}
\usepackage{epsfig}
\usepackage{url}
\usepackage{algorithm}
\usepackage{algorithmic}
\usepackage{epstopdf}
\usepackage{color}
\usepackage[tight]{subfigure}

\newcommand{\ds}{\displaystyle}
\newtheorem{myth}{Theorem}

\newtheorem{mylem}{Lemma}
\newcommand{\qed}{\mbox{}\hfill$\Box$ \vskip 3mm}

\newcommand{\la}{\langle}
\newcommand{\ra}{\rangle}
\newcommand{\Prf}{{\it Proof: \ }}

\newcommand{\clC}{{\cal C}}
\newcommand{\clL}{{\cal L}}

\newcommand{\Ex}{{\mathbb E}}
\newcommand{\st}{\mbox{s.t.}}
\newcommand{\sinr}{{\sf SINR}}

\newcommand{\clU}{{\cal U}}

\newcommand{\clR}{{\cal R}}
\newcommand{\bx}{\mathbf{x}}
\newcommand{\bA}{\mathbf{A}}
\newcommand{\bB}{\mathbf{B}}
\newcommand{\clK}{{\cal K}}
\newcommand{\clM}{{\cal M}}
\newcommand{\mbR}{\mathbb{R}}
\begin{document}
\title{Low-Latency Multiuser Two-Way Wireless Relaying  for Spectral and Energy Efficiencies}
\author{Z. Sheng, H. D. Tuan, T. Q. Duong, H. V.  Poor, and Y. Fang
\thanks{Zhichao Sheng and Hoang D. Tuan are with the school of Electrical and Data Engineering, University of Technology Sydney, Broadway, NSW 2007, Australia (email: kebon22@163.com, Tuan.Hoang@uts.edu.au)}
\thanks{Trung Q. Duong is with Queen's University Belfast, Belfast BT7 1NN, UK  (email: trung.q.duong@qub.ac.uk)}
\thanks{H. Vincent Poor is with the Department of Electrical Engineering, Princeton University, Princeton, NJ 08544, USA (e-mail: poor@princeton.edu)}
\thanks{Yong Fang is with the School of Communication and Information Engineering, Shanghai University, Shanghai, China (email: yfang@staff.shu.edu.cn)}}
\maketitle
\vspace{-1.2cm}
\begin{abstract}
The paper considers two possible approaches, which enable  multiple pairs of users to exchange information via multiple multi-antenna relays within one time-slot to save the communication bandwidth
in low-latency communications.
The first approach is to deploy full-duplexes for both users and relays to make their simultaneous
signal transmission and reception possible. In the second approach the users use a fraction of a time slot to send their information to the relays and the relays use the remaining complementary
fraction of the time slot to send the beamformed signals to the users. The inherent loop self-interference in the duplexes
and inter-full-duplexing-user interference in the first approach are absent in the second approach.
Under both these approaches,
the joint users' power allocation and relays' beamformers to either optimize the users' exchange of information or maximize the energy-efficiency  subject to user quality-of-service (QoS) in terms of the exchanging information throughput thresholds lead to complex nonconvex optimization problems. Path-following algorithms are developed for their computational solutions.
The provided numerical examples show the advantages of the second approach over the first approach.
\end{abstract}

\section{Introduction}\label{sec:intro}
Full-duplexing (FD) \cite{CM13,Hetal14,ESS14,Setal14,LM14} is a technique for simultaneous transmission and reception
 in the same time slot and over the same frequency band while two-way relaying (TWR) \cite{RW07,Amarasuriya12,Chung12,Getal13}  allows pairs
 of users to exchange their information in one step.
FD deployed at both  users and relays  thus enables the users to exchange information
via relays within a single time-slot \cite{TNT17}. This is in
contrast to the conventional one-way relaying which needs four time slots, and the half-duplexing (HD) TWR
\cite{ZhangL09,Zeng11, Chung12,STDP17}, which
needs two time slots for the same task. Thus, FD TWR seems to be a very attractive tool for
device-to-device (D2D) and machine-to-machine (M2M) communications \cite{AWM14,Liuetal15}
and low latency communication \cite{Nokia11,Jietal17,Scetal17} for Internet of Things (IoT) applications.

The major issue in FD is  the  loop self-interference (SI) due to the co-location of transmit antennas and receive antennas.
Despite considerable progress \cite{ESS14,Setal14,LM14},  it is still challenging
to attenuate the FD SI to a level such that FD can use techniques of signal processing
to outperform the conventional half-duplexing in terms of spectral and energy efficiencies \cite{TTN16,Shetal17E}.
Similarly, it is not easy to manage  TWR multi-channel interference, which becomes double  as compared
to one-way relaying  \cite{Chalise10,Phan13}. The FD-based TWR suffers even more severe interference
than the FD one-way relaying, which may reduce any throughput gain achieved by using fewer time slots \cite{TNT17}.

There is another approach to implement half-duplexing (HD) TWR within a single time slot, which avoids FD at
both users and relays. In a fraction of a time slot, the HD users send the information intended for
their partners to the relays and then
the relays  send the beamformed signals to the users within  the remaining  fraction of the time slot.
In contrast to FD relays, which use half of their available antennas for simultaneous transmission and reception,
the HD relays now can use all their antennas for separate transmissions and receptions. Thus, compared with
FD users, which need two antennas for simultaneous transmission and reception, the HD users now
need only one antenna for separate transmission and reception.

In this paper,  we consider the problem
of joint design of users' power allocation and relays' beamformers
to either maximize the user exchange information throughput or the network energy efficiency \cite{Buetal16}
subject to  user quality-of-service (QoS) constraints in terms of minimal rate thresholds. As they
constitute optimization of nonconvex objective functions subject to nonconvex constraints under both these approaches,
finding a feasible point is already challenging computationally.
Nevertheless, we develop efficient path-following algorithms for
their computation, which not only converge rapidly but also invoke a low-complexity convex quadratic
optimization problem at each iteration for generating a new and better feasible point. The numerical examples demonstrate
the full advantage of the second approach over the first approach.

The rest of this paper is organized as follows. Section II considers the two aforementioned  nonconvex problems under a
FD-based TWR setting. Section III considers them under the time-fraction (TF)-wise HD TWR setting.
Section IV verifies the full advantage of the TF-wise HD TWR over FD-based TWR via numerical examples. Section V concludes
the paper. The appendix provides some fundamental inequalities, which play a crucial role in the development
of the path-following algorithms  in the previous sections.

\emph{Notation.} Bold-faced characters denote
matrices and column vectors, with upper case used for the former and
lower case for the latter. $\pmb{X}(n,\cdot)$ represents the $n$th row of the matrix $\pmb{X}$ while  $\pmb{X}(n,m)$
is its $(n,m)$th entry. $\la \pmb{X} \ra$ is the trace of the matrix $\pmb{X}$.
 The inner product between
vectors $\pmb{x}$ and $\pmb{y}$ is defined as $\langle \pmb{x},
\pmb{y}\rangle=\pmb{x}^H\pmb{y}$.
$||.||$ is referred either to the Euclidean vector squared norm or the Frobenius matrix squared norm.
Accordingly, $||\pmb{X}||^2=\la \pmb{X}^H\pmb{X}\ra$ for any complex  $\pmb{X}$.
Lastly, $\pmb{x}\sim \mathcal{CN}(\bar{\pmb{x}},\pmb{R}_{\pmb{x}})$ means $\pmb{x}$ is a vector of Gaussian random variables with mean $\bar{\pmb{x}}$ and covariance $\pmb{R}_{\pmb{x}}$.
%

\section{Full-duplexing based two-way relaying}\label{sec:Model}
\begin{figure}[htb]
 \centering
 \centerline{\epsfig{figure=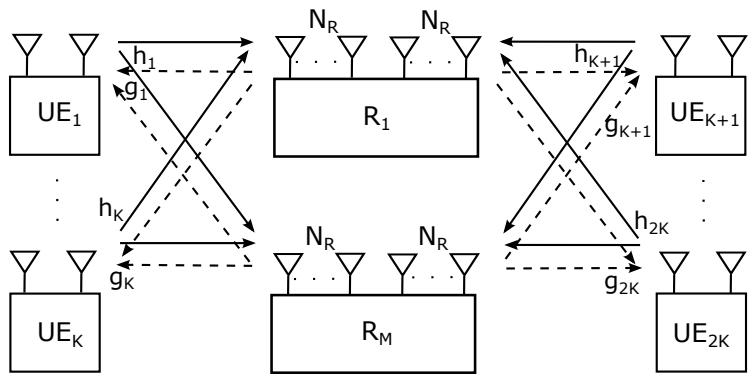,scale=1}}
\caption{Two-way relay networks with multiple two-antenna users and multiple multi-antenna relays.}
\label{fig:SystemModel}
\end{figure}

Fig. \ref{fig:SystemModel} illustrates a FD TWR network consisting of $K$ pairs of FD users (UEs) and $M$ FD relays indexed
by $m\in\clM\triangleq \{1,\dots, M\}$.
 Each FD user (UE) uses one transmit antenna and one receive antenna, while
each FD relay uses  $N_R$ receive antennas and $N_R$ transmit antennas.
Without loss of generality, the $k$th UE (UE $k$) and $(k+K)$th UE (UE $k+K$) are assumed to exchange information
with each other via the relays. The pairing operator is thus defined as $a(k)=K+k$ for $k\leq K$
and $a(k)=k-K$ if $k>K$. For each $k\in\clK\triangleq \{1,\dots, 2K\}$, define the set of UEs, which are in the same side with $k$th UE as
\[
\clU(k)=\begin{cases}\begin{array}{lll} 1, 2,...,K&\mbox{for}&1\leq k\leq K\cr
K+1,...,2K&\mbox{for}&k\geq K+1.
\end{array}\end{cases}
\]
Under simultaneous transmission and reception,  FD UEs in $\clU(k)$ interfere each other. Such kind of interference is called inter-FD-user interference.

Let $\pmb{s}=[s_1,\ldots,s_{2K}] \in \clC^{2K}$ be the vectors of information symbols $s_k$ transmitted
from UEs, which are independent and have unit energy, i.e.  $\Ex[\pmb{s}\pmb{s}^H]=\pmb{I}_{2K}$.
For $\pmb{h}_{\ell,m}\in \mathbb{C}^{N_R}$ as the vector of channels from UE $\ell$ to  relay $m$,
the received signal at  relay $m$ is
\begin{equation}\label{receivedr}
\pmb{r}_m=\sum_{\ell\in\clK}\sqrt{p_{\ell}} \pmb{h}_{\ell,m} s_{\ell}+e_{LI,m}+\pmb{n}_{R,m},
\end{equation}
where $\pmb{n}_{R,m}\sim \mathcal{CN}(0,\sigma_R^2\pmb{I}_{N_R})$ is the background noise,
and $\pmb{p}=(p_1,\ldots,p_{2K})$ is a vector of UE power allocation, while
$e_{LI,m}\in\mathbb{C}^{N_R}$ models the effect of analog circuit non-ideality and the limited dynamic range of the
analog-to-digital converter (ADC) at FD relay $m$.

The transmit power at UEs is physically limited  by $P^{U,\max}$ as
\begin{equation}\label{IndiUserCon}
p_k \leq P^{U,\max}, \quad k\in\clK.
\end{equation}
The total transmit power of UEs is bounded by $P_{\mathrm{sum}}^{U,\max}$ to prevent their excessive interference to other networks as
\begin{equation}\label{SumUserCon}
P_{\mathrm{sum}}^{U}(\pmb{p})=\sum_{k\in\clK} p_k \leq P_{\mathrm{sum}}^{U,\max}.
\end{equation}
Relay $m$ processes the received signal by applying the beamforming matrix
$\pmb{W}_m \in \clC^{N_R \times N_R}$ for transmission:
\begin{equation}
\pmb{r}_{m,b}=\pmb{W}_m\pmb{r}_m=\sum_{\ell\in\clK} \sqrt{p_{\ell}} \pmb{W}_m\pmb{h}_{\ell,m} s_{\ell}+\pmb{W}_m(e_{LI,m}+\pmb{n}_{R,m}).
\end{equation}
For simplicity it is assumed that $\pmb{W}_me_{LI,m}\sim \mathcal{CN}(0,\sigma_{SI}^2P_{m}^A(\pmb{p},\pmb{W}_m)I_{N_R})$ with the relay channel's instantaneous residual SI attenuation level $\sigma_{SI}$.\footnote{It is more practical to assume $e_{LI,m}\sim \mathcal{CN}(0,\bar{\sigma}_{SI}^2P_{m}^A(\pmb{p},\pmb{W}_m)I_{N_R})$
so $\pmb{W}_me_{LI,m}\sim \mathcal{CN}(0,\bar{\sigma}_{SI}^2P_{m}^A(\pmb{p},\pmb{W}_m)\pmb{W}_{m}\pmb{W}_m^H)$
resulting in
$\mathbb{E}[||\pmb{W}_me_{LI,m}||^2]=\bar{\sigma}_{SI}^2P_{m}^A(\pmb{p},\pmb{W}_m)||\pmb{W}_m||^2$. Usually $||\pmb{W}_m||^2\leq
\nu$ can be assumed so $\mathbb{E}[||\pmb{W}_me_{LI,m}||^2]=\sigma_{SI}^2P_{m}^A(\pmb{p},\pmb{W}_m)$ for $\sigma_{SI}^2=\nu\bar{\sigma}_{SI}^2$ } This gives
\[
\mathbb{E}[||\pmb{W}_me_{LI,m}||^2]=\sigma_{SI}^2P_{m}^A(\pmb{p},\pmb{W}_m),
\]
in calculating the transmit power at relay $m$  by  a closed-form as
\begin{eqnarray}
  P_{m}^A(\pmb{p},\pmb{W}_m) & =&\Ex[||\pmb{r}_{m,b}||^2] \nonumber \\
   & =& \ds \sum_{\ell\in\clK} p_{\ell} ||\pmb{W}_m\pmb{h}_{\ell,m}||^2+
   \sigma_R^2||\pmb{W}_m||^2+\mathbb{E}[||\pmb{W}_me_{LI,m}||^2]\nonumber\\
   &=&[\ds \sum_{\ell\in\clK} p_{\ell}||\pmb{W}_m\pmb{h}_{\ell,m}||^2+
   \sigma_R^2||\pmb{W}_m||^2]/(1-\sigma_{SI}^2).\label{rpower}
\end{eqnarray}
This transmit power at  relay $m$ must be physically limited by a  physical parameter $P^{A,\max}$ as
\begin{equation}\label{IndiRelayCon}
 P_{m}^A(\pmb{p},\pmb{W}_m)\leq P^{A,\max}, m\in\clM,
\end{equation}
and their sum is also bounded by $P_{\mathrm{sum}}^{R,\max}$ to control the network emission to other networks:
\begin{equation}\label{SumRelayCon}
\begin{array}{ll}
  P_{\mathrm{sum}}^R(\pmb{p},\pmb{W}) & =\ds \sum_{m\in\clM} P_{m}^A(\pmb{p},\pmb{W}_m) \\
   & =\ds\sum_{m\in\clM}[\ds \sum_{k\in\clK} p_{\ell}||\pmb{W}_m\pmb{h}_{\ell,m}||^2+
   \sigma_R^2||\pmb{W}_m||^2]/(1-\sigma_{SI}^2)\leq P_{\mathrm{sum}}^{R,\max}.
\end{array}
\end{equation}
The relays transmit the processed signals to all UEs.
For  the vector channel $\pmb{g}_{m,k}\in \clC^{N_R}$ from relay $m$ to UE $k$ and
channel $\chi_{\eta,k}$ from UE $\eta\in\clU(k)$ to UE $k$, the received signal at UE $k$ is given by
\begin{eqnarray}
y_k&=&\ds\sum_{m\in\clM}\pmb{g}_{m,k}^T\pmb{r}_{m,b}+{\color{black}
\sum_{\eta\in\clU(k)}\chi_{\eta,k}\sqrt{p}_{\eta}\tilde{s}_{\eta}}+n_k\nonumber\\
&=&\ds\sum_{m\in\clM}\pmb{g}_{m,k}^T
[\sum_{\ell\in\clK} \sqrt{p_\ell}\pmb{W}_m\pmb{h}_{\ell,m} s_\ell+\pmb{W}_m(e_{LI,m}+\pmb{n}_{R,m})]+
{\color{black}
\sum_{\eta\in\clU(k)}\chi_{\eta,k}\sqrt{p}_{\eta}\tilde{s}_{\eta}}
+n_k,\label{receivedu}
\end{eqnarray}
where  $n_{k}\sim \mathcal{CN}(0,\sigma_{k}^2)$ is the background noise,
and $|\chi_{k,k}|^2=\sigma_{SI}^2$ as
$\chi_{k,k}\tilde{s}_k$ represents the loop interference at UE $k$.
We can rewrite (\ref{receivedu}) as
\begin{eqnarray}
y_k&=&\ds\sqrt{p_{a(k)}}\sum_{m\in\clM}\pmb{g}_{m,k}^T
\pmb{W}_m\pmb{h}_{a(k),m} s_{a(k)} +\sqrt{p_k}\sum_{m\in\clM}\pmb{g}_{m,k}^T
\pmb{W}_m\pmb{h}_{k,m} s_{k}\nonumber\\
&&+ \ds\sum_{m\in\clM}\pmb{g}_{m,k}^T\left[\sum_{\ell\in\clK\setminus\{k, {a(k)}\}}\sqrt{p_\ell}\pmb{W}_m\pmb{h}_{\ell,m} s_\ell+\pmb{W}_m(e_{LI,m}+\pmb{n}_{R,m})\right]\nonumber\\
&&+\ds\sum_{\eta\in\clU(k)}\chi_{\eta,k}\sqrt{p}_{\eta}\tilde{s}_{\eta}+n_k.\label{yk}
\end{eqnarray}
Note that the first term in \eqref{yk} is the desired signal component,  the third term is the inter-pair interference and the last two terms are noise. UE $k$ can cancel the self-interference by the second term using
the channel state information of the forward channels $\mathbf{h}_{k,m}$ from itself to the relays
and backward channels $\mathbf{g}_{m,k}$ from the relays to itself as
 well as the beamforming matrix $\mathbf{W}_m$. The challenges here is that the loop SI term $
\sum_{\eta\in\clU(k)}\chi_{\eta,k}\sqrt{p}_{\eta}\tilde{s}_{\eta}$, which may be strong due to the proximity of
UEs in  ${\cal U}(k)$, cannot be nulled out. This means more power should be given to the relays  but it
leads to more FD SI at the relays.

Furthermore, for $\pmb{f}_{m,k}\triangleq\pmb{g}_{m,k}^*$,
the signal-to-interference-plus-noise ratio ($\sinr$) at UE $k$' receiver can be calculated as
\begin{eqnarray}
\gamma_{k}(\pmb{p}, \pmb{W})&=&\ds {\color{black}p_{a(k)}\ds\left|\sum_{m\in\clM}\pmb{f}_{m,k}^H
\pmb{W}_m\pmb{h}_{a(k),m}\right|^2}/
\left[\ds\sum_{\ell\in\clK\setminus\{k,a(k)\}} p_{\ell}\left|\sum_{m\in\clM}\pmb{f}_{m,k}^H \pmb{W}_m\pmb{h}_{\ell,m}\right|^2\right.\nonumber\\
&&+\sigma_R^2\sum_{m\in\clM}||\pmb{f}_{m,k}^H \pmb{W}_m||^2+\frac{\sigma_{SI}^2}{1-\sigma_{SI}^2}
\sum_{m\in\clM}||\pmb{g}_{m,k}||^2(\ds \sum_{\ell\in\clK} p_{\ell}||\pmb{W}_m\pmb{h}_{\ell,m}||^2\nonumber\\
&&\left.+   \sigma_R^2||\pmb{W}_m||^2)+{\color{black}
\sum_{\eta\in\clU(k)}|\chi_{\eta,k}|^2p_{\eta}} +\sigma^2_{k}\right].
\label{SINR}
\end{eqnarray}
Under the definitions
\begin{equation}\label{ndef1}
\begin{array}{c}
\clL_{k,\ell}(\pmb{W})\triangleq \ds{\color{black}\sum_{m\in\clM}\pmb{f}_{m,k}^H \pmb{W}_m\pmb{h}_{\ell,m}},\\
\clL_k(\pmb{W})\triangleq\begin{bmatrix}\pmb{f}_{1,k}^H \pmb{W}_1&\pmb{f}_{2,k}^H \pmb{W}_2&...&\pmb{f}_{M,k}^H \pmb{W}_M
\end{bmatrix},
\end{array}
\end{equation}
it follows that
\begin{eqnarray}
\gamma_{k}(\pmb{p}, \pmb{W})&=&\ds p_{a(k)}{\color{black}|\clL_{k,a(k)}(\pmb{W})|^2}/
\ds\left[\sum_{\ell\in\clK\setminus\{k,a(k)\}} p_{\ell}{\color{black}|\clL_{k,\ell}(\pmb{W})|^2}\right.\nonumber\\
&&+\sigma_R^2||\clL_k(\pmb{W})||^2+\frac{\sigma_{SI}^2}{1-\sigma_{SI}^2}
\sum_{m\in\clM}||\pmb{g}_{m,k}||^2(\ds \sum_{\ell\in\clK} p_{\ell}||\pmb{W}_m\pmb{h}_{\ell,m}||^2\nonumber\\
&&\ds\left.+   \sigma_R^2||\pmb{W}_m||^2)
+{\color{black}
\sum_{\eta\in\clU(k)}|\chi_{\eta,k}|^2p_{\eta}} +\sigma^2_{k}\right].
\label{SINR.e}
\end{eqnarray}
In FD TWR, the performance of interest is the exchange information throughput of UE pairs:
\begin{eqnarray}
R_k(\pmb{p}, \pmb{W})&=& \ln ( 1+\gamma_{k}(\pmb{p}, \pmb{W}))+\ln (1+\gamma_{a(k)}(\pmb{p}, \pmb{W})),  k=1, \dots, K.\label{Ratek}
\end{eqnarray}
The problem  of  maximin exchange information throughput optimization subject to transmit power constraints
is then formulated as
\begin{subequations}\label{MaxMinPair2}
\begin{eqnarray}
&\ds\max_{\pmb{W}, \pmb{p}} & \ds \min_{k=1,\ldots ,K}\ \left[\ln (1+ \gamma_{k}(\pmb{p}, \pmb{W}))+\ln (1+\gamma_{a(k)}(\pmb{p}, \pmb{W}))\right] \label{MaxMinPairObj2}\\
& \st & \eqref{IndiUserCon}, \eqref{SumUserCon}, \eqref{IndiRelayCon}, \eqref{SumRelayCon}.
\end{eqnarray}
\end{subequations}
Another problem, which attracted recent attention in 5G \cite{Buetal16,Zaetal16}
is the following problem of maximizing the network energy-efficiency (EE)
subject to UE QoS in terms of the exchange information throughput thresholds:
\begin{subequations}\label{e1}
\begin{eqnarray}
&\ds\max_{\pmb{W}, \pmb{p}} & \ds\frac{ \ds\sum_{k=1}^K\left[\ln (1+ \gamma_{k}(\pmb{p}, \pmb{W}))+\ln (1+\gamma_{a(k)}(\pmb{p}, \pmb{W}))\right]}{\zeta(P^U_{\rm sum}(\mathbf{p})+P^R_{\rm sum}(\mathbf{p},
\mathbf{W}))+MP^{\rm R}+2KP^{\rm U}} \label{e1a}\\
& \st & \eqref{IndiUserCon}, \eqref{SumUserCon}, \eqref{IndiRelayCon}, \eqref{SumRelayCon},\label{e1b}\\
&&R_k(\pmb{p}, \pmb{W})\geq r_k, k=1,\dots, K,\label{e1c}
\end{eqnarray}
\end{subequations}
where $\zeta$, $P^{\rm R}$ and $P^{\rm U}$ are the reciprocal of drain efficiency of power amplifier,
the circuit powers of the relay and UE, respectively, and $r_k$ sets the exchange throughput threshold for UE pairs.

The next two subsections are devoted to computational solution for problems (\ref{MaxMinPair2}) and (\ref{e1}),
respectively.
\subsection{FD TWR maximin exchange information throughput optimization}
By introducing new nonnegative variables
\begin{equation}\label{betak}
\beta_k=1/p_k^2 > 0, k\in\clK,
\end{equation}
and functions
\begin{equation}\label{Psi}
\begin{array}{c}
 \Psi_{k,\ell}(\pmb{W}, \alpha, \beta)\triangleq |\clL_{k,\ell}(\pmb{W})|^2/\sqrt{\alpha\beta}, (k,\ell)\in\clK\times\clK,\\
\Upsilon_{k}(\pmb{W}, \alpha)\triangleq ||\clL_k(\pmb{W})||^2/\sqrt{\alpha}, k\in\clK,\\
\Phi_{\ell,m}(\pmb{W}_m, \alpha,\beta)\triangleq \ds ||\pmb{h}_{\ell,m}^H\pmb{W}_m||^2/\sqrt{\alpha\beta}, \quad (\ell,m)\in\clK\times\clM,
\end{array}
\end{equation}
which are convex \cite{DM08}, (\ref{SINR.e}) can be re-expressed by
\begin{eqnarray}
\gamma_{k}(\pmb{p}, \pmb{W})&=&\ds {\color{black} |\clL_{k,a(k)}(\pmb{W})|^2}/\sqrt{\beta_{a(k)}}
\left[\ds\sum_{\ell\in\clK\setminus\{k,a(k)\}}\Psi_{k,\ell}(\pmb{W},1,\beta_{\ell})\right.\nonumber\\
&&+\sigma_R^2\Upsilon_k(\pmb{W},1)+\frac{\sigma_{SI}^2}{1-\sigma_{SI}^2}
\sum_{m\in\clM}||\pmb{g}_{m,k}||^2\ds\left( \sum_{\ell\in\clK} \Phi_{\ell,m}(\pmb{W}_m,1,\beta_{\ell})\right.\nonumber\\
&&\ds\left.+   \sigma_R^2\la \pmb{W}_m^H\pmb{W}_m\ra\right)
\ds\left.+{\color{black}
\sum_{\eta\in\clU(k)}|\chi_{\eta,k}|^2/\sqrt{\beta_{\eta}}} +\sigma^2_{k}\right].
\label{SINR.ea}
\end{eqnarray}
Similarly to \cite{Khetal13} and \cite[Th. 1]{STDP17} we can prove the following result.
\begin{myth}\label{basic}
The optimization problem \eqref{MaxMinPair2}, which is maximization of nonconcave objective
function over a nonconvex set,
can be equivalently rewritten as the following problem of maximizing a nonconcave objective function over a  set
of convex constraints:
\allowdisplaybreaks[4]
\begin{subequations}\label{MaxMinPair3}
\begin{eqnarray}
&\ds\max_{\pmb{W},\pmb{\alpha}, \pmb{\beta}} f(\pmb{W},\pmb{\alpha},
\pmb{\beta})\triangleq
\ds \min_{k=1,\ldots ,K}\
\left[\ln (1+|\clL_{k,a(k)}(\pmb{W})|^2/
\sqrt{\alpha_{k}\beta_{a(k)}})\right.\nonumber\\
&\left.\hspace*{5cm}+\ln (1+|\clL_{a(k),k}(\pmb{W})|^2/
\sqrt{\alpha_{a(k)}\beta_{k}}) \right] & \label{MaxMinPairObj3}\\
& \st \quad \ds \sum_{\ell\in\clK\setminus\{k,a(k)\}} \Psi_{k,\ell}(\pmb{W}, \alpha_k, \beta_{\ell})
+\sigma_R^2 \Upsilon_k(\pmb{W},\alpha_k) +
{\color{black}
\sum_{\eta\in\clU(k)}|\chi_{\eta,k}|^2/\sqrt{\alpha_k\beta_{\eta}}}&\nonumber\\
&+\ds\frac{\sigma_{SI}^2}{1-\sigma_{SI}^2}
\sum_{m\in\clM}||\pmb{g}_{m,k}||^2(\ds \sum_{\ell\in\clK}\Phi_{\ell,m}(\pmb{W}_m,\alpha_k,\beta_{\ell}) +
 \sigma_R^2 ||\pmb{W}_m||^2/\sqrt{\alpha_k})+\sigma^2_{k}/\sqrt{\alpha_k}  \leq 1,& \label{DemoniCon3}\\
& \beta_k \geq 1/(P^{U,\max})^2, \hspace*{7cm} k\in\clK, & \label{UserCon3}\\
& P_{\mathrm{sum}}^{U}(\pmb{\beta}):=\ds \sum_{k\in\clK}1/\sqrt{\beta_k} \leq P_{\mathrm{sum}}^{U,\max} & \label{TotalUserCon3}\\
& \ds \sum_{\ell\in\clK}\Phi_{\ell,m}(\pmb{W}_m,1,\beta_{\ell}) +\sigma_R^2||\pmb{W}_m||^2  \leq (1-\sigma^2_{SI})P_m^{A,\max},  m\in\clM,  \label{RelayCon3}\\
& \ds\sum_{m\in\clM}\left[\ds \sum_{\ell\in\clK}\Phi_{\ell,m}(\pmb{W}_m,1,\beta_{\ell}) +\sigma_R^2||\pmb{W}_m||^2 \right]
\leq (1-\sigma^2_{SI})P_{\mathrm{sum}}^{R,\max}. & \label{TotalRelayCon3}
\end{eqnarray}
\end{subequations}
\end{myth}
The main issue now is to handle the nonconcave objective function in (\ref{MaxMinPairObj3}) of (\ref{MaxMinPair3}),
which is resolved by the following theorem.
\begin{myth}\label{cth1} At any $(\pmb{W}^{(\kappa)},\pmb{\alpha}^{(\kappa)},\pmb{\beta}^{(\kappa)})$ feasible for
the convex constraints (\ref{DemoniCon3})-(\ref{TotalRelayCon3}) it  is true that
\begin{eqnarray}
\ln (1+|\clL_{k,a(k)}(\pmb{W})|^2/
\sqrt{\alpha_{k}\beta_{a(k)}})
&\geq&f^{(\kappa)}_{k,a(k)}(\pmb{W},\alpha_{k},\beta_{a(k)})\label{ap1}
\end{eqnarray}
over the trust region{\color{black}
\begin{equation}\label{tru1}
2\Re\{\clL_{k,a(k)}(\pmb{W})(\clL_{k,a(k)}(\pmb{W}^{(\kappa)}))^*\}
-|\clL_{k,a(k)}(\pmb{W}^{(\kappa)})|^2>0,
\end{equation}}
for
\begin{eqnarray}\label{ap1a}
f^{(\kappa)}_{k,a(k)}(\pmb{W},\alpha_{k},\beta_{a(k)})&=&\nonumber\\
\ln(1+x_{k,a(k)}^{(\kappa)})+
a_{k,a(k)}^{(\kappa)}[2-\ds\frac{|\clL_{k,a(k)}(\pmb{W}^{(\kappa)})|^2}{{\color{black} 2\Re\{\clL_{k,a(k)}(\pmb{W})(\clL_{k,a(k)}(\pmb{W}^{(\kappa)}))^*\}}
-|\clL_{k,a(k)}(\pmb{W}^{(\kappa)})|^2}&&\nonumber\\
-\sqrt{\alpha_k\beta_{a(k)}}/\sqrt{\alpha^{(\kappa)}_{k}\beta^{(\kappa)}_{a(k)}}&&
\end{eqnarray}
with $x_{k,a(k)}^{(\kappa)}\triangleq|\clL_{k,a(k)}(\pmb{W}^{(\kappa)})|^2/\sqrt{\alpha^{(\kappa)}_{k}\beta^{(\kappa)}_{a(k)}}$
and $a_{k,a(k)}^{(\kappa)}\triangleq x_{k,a(k)}^{(\kappa)})/(x_{k,a(k)}^{(\kappa)}+1)>0$.
\end{myth}
\Prf (\ref{ap1a}) follows by applying inequality (\ref{ine1p}) in the Appendix for
\[
x=1/|\clL_{k,a(k)}(\pmb{W})|^2, y=\sqrt{\alpha_{k}\beta_{a(k)}}
\]
and
\[
\bar{x}=1/|\clL_{k,a(k)}(\pmb{W}^{(\kappa)})|^2, \bar{y}=\sqrt{\alpha^{(\kappa)}_{k}\beta^{(\kappa)}_{a(k)}}
\]
and then the inequality
\begin{equation}\label{inequa1}
1/|\clL_{k,a(k)}(\pmb{W})|^2\leq 1/\left(2\Re\{\clL_{k,a(k)}(\pmb{W})(\clL_{k,a(k)}(\pmb{W}^{(\kappa)}))^*\}
-|\clL_{k,a(k)}(\pmb{W}^{(\kappa)})|^2\right)
\end{equation}
over the trust region (\ref{tru1}).\qed

By Algorithm \ref{alg1} we propose a path-following procedure for computing (\ref{MaxMinPair3}), which
solves the following convex optimization problem of inner approximation
at the $\kappa$th iteration to generate the next feasible  point $(\pmb{W}^{(\kappa+1)}, \pmb{\alpha}^{(\kappa+1)}, \pmb{\beta}^{(\kappa+1)})$:
\begin{equation}\label{ConvexOpt}
\ds\max_{\pmb{W}, \pmb{\alpha}, \pmb{\beta}} \ds \min_{k=1,\ldots ,K}
[f^{(\kappa)}_{k,a(k)}(\pmb{W},\alpha_{k},\beta_{a(k)})+f^{(\kappa)}_{a(k),k}(\pmb{W},\alpha_{a(k)},\beta_{k})]\quad
\mbox{s.t.}\quad (\ref{DemoniCon3})-(\ref{TotalRelayCon3} ), (\ref{tru1}).
\end{equation}
Similarly to \cite[Alg. 1]{STDP17}, it can be shown that the sequence $\{ (\pmb{W}^{(\kappa)},\pmb{\alpha}^{(\kappa)},\pmb{\beta}^{(\kappa)}) \}$ generated by Algorithm \ref{alg1}
at least converges to {\color{black}a local optimal solution}  of  (\ref{MaxMinPair3}).\footnote{As mentioned
in \cite[Remark]{MW78} this desired property of a limit point indeed does not require the differentiability of the objective function }
\begin{algorithm}
\caption{Path-following algorithm for FD TWR exchange throughput optimization} \label{alg1}
\begin{algorithmic}
\STATE \textbf{initialization}: Set $\kappa=0$. Initialize a feasible point
$(\pmb{W}^{(0)}, \pmb{\alpha}^{(0)}, \pmb{\beta}^{(0)})$ for the convex constraints
(\ref{DemoniCon3})-(\ref{TotalRelayCon3}) and
 $R_1=f(\pmb{W}^{(0)}, \pmb{\alpha}^{(0)}, \pmb{\beta}^{(0)})$.
\REPEAT \STATE $\bullet$ $R_0= R_1$.
\STATE $\bullet$ Solve the
convex optimization problem \eqref{ConvexOpt} to obtain the
solution $(\pmb{W}^{(\kappa+1)}, \pmb{\alpha}^{(\kappa+1)}, \pmb{\beta}^{(\kappa+1)})$.
\STATE $\bullet$ Update $R_1=f(\pmb{W}^{(\kappa+1)}, \pmb{\alpha}^{(\kappa+1)}, \pmb{\beta}^{(\kappa+1)})$.
\STATE $\bullet$ Reset $\kappa \to\kappa+1$.
 \UNTIL{$\frac{R_1-R_0}{R_0} \leq
 \epsilon$ for given tolerance $\epsilon>0$}.
\end{algorithmic}
\end{algorithm}
\subsection{FD TWR energy-efficiency maximization}
We return to consider the optimization problem (\ref{e1}), which can be shown
similarly to Theorem \ref{basic} to be equivalent to  the following optimization
problem under the variable change (\ref{betak}):
\begin{subequations}\label{e1.eq}
\begin{eqnarray}
\ds\max_{\pmb{W}, \pmb{\alpha}, \pmb{\beta}}\ F(\pmb{W},\pmb{\alpha},
\pmb{\beta})  \quad
\mbox{s.t.} \quad (\ref{DemoniCon3})-(\ref{TotalRelayCon3}),\label{eq.eqa}\\
\tilde{R}_k(\mathbf{W},\pmb{\alpha},\pmb{\beta})\geq r_k, k=1,\dots, K,\label{e1.eqc}
\end{eqnarray}
\end{subequations}
for
\[
\begin{array}{c}
F(\pmb{W},\pmb{\alpha},
\pmb{\beta})\triangleq\ds\left[\sum_{k=1}^K
\tilde{R}_k(\mathbf{W},\pmb{\alpha},\pmb{\beta})\right]/\pi(\pmb{\beta},\pmb{W}),\\
\tilde{R}_k(\mathbf{W},\pmb{\alpha},\pmb{\beta})\triangleq \ds\ln \left(1+|\clL_{k,a(k)}(\pmb{W})|^2/
\sqrt{\alpha_{k}\beta_{a(k)}}\right)+\ln \left(1+|\clL_{a(k),k}(\pmb{W})|^2/
\sqrt{\alpha_{a(k)}\beta_{k}}\right),
\end{array}
\]
and
\begin{eqnarray}\label{e2}
\pi(\pmb{\beta},\pmb{W})&\triangleq & \ds\sum_{k\in\clK}\zeta/\sqrt{\beta_k}+(\zeta/(1-\sigma_{SI}^2))\ds\sum_{m\in\clM}
\left[\ds \sum_{\ell\in\clK}\Phi_{\ell,m}(\pmb{W}_m,1,\beta_{\ell})\right.\nonumber\\
&&\ds\left. +\sigma_R^2||\pmb{W}_m||^2 \right]+MP^{\rm R}+2KP^{\rm U}.
\end{eqnarray}
The objective function in (\ref{eq.eqa}) is nonconcave and constraint (\ref{e1.eqc}) is nonconvex.

Suppose that $(\pmb{W}^{(\kappa)},\pmb{\alpha}^{(\kappa)},\pmb{\beta}^{(\kappa)})$ is a feasible point for (\ref{e1.eq})
found from the $(\kappa-1)$th iteration.
Applying inequality (\ref{ine1}) in the Appendix for
\[
x=1/|\clL_{k,a(k)}(\pmb{W})|^2, y=\sqrt{\alpha_{k}\beta_{a(k)}},
t=\pi(\pmb{\beta},\pmb{W})
\]
and
\[
\bar{x}=1/|\clL_{k,a(k)}(\pmb{W}^{(\kappa)})|^2, \bar{y}=\sqrt{\alpha_{k}^{(\kappa)}\beta_{a(k)}^{(\kappa)}},
\bar{t}=\pi(\pmb{\beta}^{(\kappa)},\pmb{W}^{(\kappa)})
\]
and using inequality (\ref{inequa1}) yield the following  bound for the terms of the objective function in (\ref{eq.eqa}):
\begin{eqnarray}
\ds\left[\ln (1+|\clL_{k,a(k)}(\pmb{W})|^2/
\sqrt{\alpha_{k}\beta_{a(k)}})\right]/\pi(\pmb{\beta},\pmb{W})
&\geq&F_{k,a(k)}^{(\kappa)}(\pmb{W},\alpha_{k},
\pmb{\beta})\label{e7}
\end{eqnarray}
over the trust region (\ref{tru1}),  where
\[
\allowdisplaybreaks[4]
\begin{array}{lll}
F_{k,a(k)}^{(\kappa)}(\pmb{W},\alpha_{k},
\pmb{\beta})&\triangleq&\ds p_{k,a(k)}^{(\kappa)}+q_{k,a(k)}^{(\kappa)}
\ds\left[2-\frac{|\clL_{k,a(k)}(\pmb{W}^{(\kappa)})|^2}
{{\color{black}2\Re\{\clL_{k,a(k)}(\pmb{W})(\clL_{k,a(k)}(\pmb{W}^{(\kappa)}))^*\}}
-|\clL_{k,a(k)}(\pmb{W}^{(\kappa)})|^2}\right.\nonumber\\[0.2cm]
&&-\ds\left.\sqrt{\alpha_k\beta_{a(k)}}/\sqrt{\alpha^{(\kappa)}_{k}\beta^{(\kappa)}_{a(k)}}
\right]-r_{k,a(k)}^{(\kappa)}\pi(\pmb{\beta},\pmb{W}),
\end{array}
\]
and
\begin{equation}\label{e8}
\begin{array}{lll}
x_{k,a(k)}^{(\kappa)}&=&|\clL_{k,a(k)}(\pmb{W}^{(\kappa)})|^2/\sqrt{\alpha^{(\kappa)}_{k}\beta^{(\kappa)}_{a(k)}},\\
t^{(\kappa)}&=&\pi(\pmb{\beta}^{(\kappa)},\pmb{W}^{(\kappa)}),\\
p_{k,a(k)}^{(\kappa)}&=&2\ds \left[\ln(1+x_{k,a(k)}^{(\kappa)})\right]/t^{(\kappa)}>0,\\
q_{k,a(k)}^{(\kappa)}&=&\ds x_{k,a(k)}^{(\kappa)}/((x_{k,a(k)}^{(\kappa)}+1)t^{(\kappa)})>0, \\
r_{k,a(k)}^{(\kappa)}&=&\ds\left[\ln(1+x_{k,a(k)}^{(\kappa)})\right]/(t^{(\kappa)})^2>0.
\end{array}
\end{equation}
Furthermore, we use $f^{(\kappa)}_{k,a(k)}$  defined from (\ref{ap1}) to provide the following inner convex approximation
for the nonconvex constraint (\ref{e1.eqc}):
\begin{equation} f^{(\kappa)}_{k,a(k)}(\pmb{W},\alpha_{k},\beta_{a(k)}) +f^{(\kappa)}_{a(k),k}(\pmb{W},\alpha_{a(k)},\beta_{k})
\geq r_k.\label{opt.c}
\end{equation}
By Algorithm \ref{alg2} we propose a path-following procedure for computing (\ref{e1.eq}), which
solves the following convex optimization problem at the $\kappa$th iteration to generate the next feasible  point $(\pmb{W}^{(\kappa+1)}, \pmb{\alpha}^{(\kappa+1)}, \pmb{\beta}^{(\kappa+1)})$:
\begin{subequations}\label{ConvexOpt.e}
\begin{eqnarray}
\ds\max_{\pmb{W}, \pmb{\alpha}, \pmb{\beta}} F(\pmb{W},\pmb{\alpha},
\pmb{\beta})\triangleq
\ds {\color{black}
\sum_{k=1}^K}
[F^{(\kappa)}_{k,a(k)}(\pmb{W},\alpha_{k},\pmb{\beta})+F^{(\kappa)}_{a(k),k}(\pmb{W},\alpha_{a(k)},\pmb{\beta})]\label{opt.a}\\
\mbox{s.t.}\quad (\ref{DemoniCon3})-(\ref{TotalRelayCon3} ), (\ref{tru1}), (\ref{opt.c}) \label{opt.b}
\end{eqnarray}
\end{subequations}
Analogously to Algorithm \ref{alg1}, the sequence $\{ (\pmb{W}^{(\kappa)},\pmb{\alpha}^{(\kappa)},\pmb{\beta}^{(\kappa)}) \}$ generated by Algorithm \ref{alg2}
at least converges to a local optimal solution  of  (\ref{e1.eq}).

An initial feasible point $(\pmb{W}^{(0)}, \pmb{\alpha}^{(0)}, \pmb{\beta}^{(0)})$ for initializing
Algorithm \ref{alg2} can be found by using Algorithm \ref{alg1} for
computing (\ref{MaxMinPair2}), which terminates upon
\begin{equation}\label{inipoint}
\ds{\min_{k=1,\ldots ,K}} R_k(\pmb{W}^{(\kappa)}, \pmb{\alpha}^{(\kappa)}, \pmb{\beta}^{(\kappa)})/r_k\geq 1
\end{equation}
to satisfy (\ref{e1.eqc}).
\begin{algorithm}
\caption{Path-following algorithm for FD TWR energy-efficiency} \label{alg2}
\begin{algorithmic}
\STATE \textbf{initialization}: Set $\kappa=0$. Initialize a feasible point
$(\pmb{W}^{(0)}, \pmb{\alpha}^{(0)}, \pmb{\beta}^{(0)})$ for (\ref{e1.eq}) and
$e_1=F(\pmb{W}^{(0)}, \pmb{\alpha}^{(0)}, \pmb{\beta}^{(0)})$.
\REPEAT \STATE $\bullet$ $e_0= e_1$.
\STATE $\bullet$ Solve the
convex optimization problem \eqref{ConvexOpt.e} to obtain the
solution $(\pmb{W}^{(\kappa+1)}, \pmb{\alpha}^{(\kappa+1)}, \pmb{\beta}^{(\kappa+1)})$.
\STATE $\bullet$ Update $e_1=F(\pmb{W}^{(\kappa+1)}, \pmb{\alpha}^{(\kappa+1)}, \pmb{\beta}^{(\kappa+1)})$.
\STATE $\bullet$ Reset $\kappa \to\kappa+1$.
 \UNTIL{$\frac{e_1-e_0}{e_0} \leq
 \epsilon$ for given tolerance $\epsilon>0$}.
\end{algorithmic}
\end{algorithm}
\section{Time-fraction-wise HD two-way relaying}
\begin{figure}[htb]
 \centering
 \centerline{\epsfig{figure=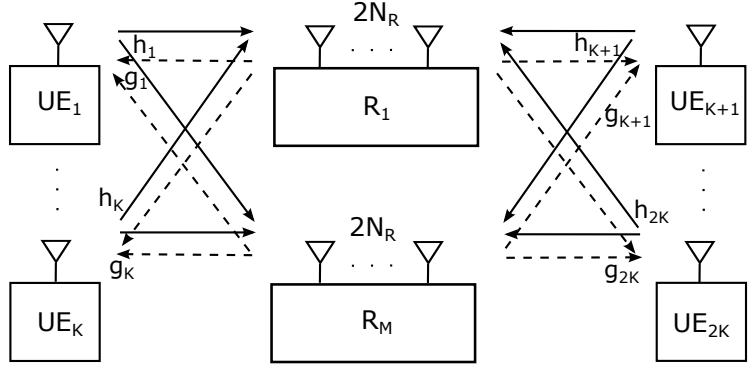,scale=1}}
\caption{Two-way relay networks with multiple single-antenna users and multiple multi-antenna relays.}
\label{fig:SystemModel_FT}
\end{figure}

Through the FD-based TWR detailed in the previous section one can see the following obvious issues for its practical
implementations:
\begin{itemize}
\item It is difficult to attenuate FD SI at the UEs and relays to a level in realizing the benefits by FD.
The FD SI is even more severe at the relays, which are equipped with multiple antennas;
\item Inter-FD-user  interference cannot be controlled;
\item It is technically difficult to implement FD at UEs, which particularly requires two antennas per UE.
\end{itemize}
We now propose a new way for UE information exchange via HD TWR within the time slot as illustrated by
Fig. \ref{fig:SystemModel_FT}, where at time-fraction $0<\tau<1$ all UEs send information to the relays and at the remaining time fraction $(1-\tau)$ the relays send the beamformed signals to UEs. This alternative
has the following advantages:
\begin{itemize}
\item Each relay uses all available $2N_R$ antennas for separated receiving and transmitting signals;
\item UEs need only a single antenna to implement the conventional HD, which transmits signal
and receive signals in separated time fractions.
\end{itemize}

Suppose that UE $k$ uses the power {\color{black}$\tau p_k$} to send information to the delay. The following
physical limitation is imposed:
\begin{equation}\label{al1}
{\color{black}p_k\leq \bar{P}_{\rm UE}}, k\in\clK,
\end{equation}
where {\color{black}$\bar{P}_{\rm UE}$ is a physical parameter to signify the hardware limit
in transmission during time-fractions. Typically, $\bar{P}_{\rm UE}=3P^{U,\max}$ for
$P^{U,\max}$ defined from (\ref{IndiUserCon}).}

As in (\ref{SumUserCon}), the  power budget of all UEs is  $P_{\mathrm{sum}}^{U,\max}$:
\begin{equation}\label{al2}
P_{\mathrm{sum}}^{U}(\pmb{p})=\tau \sum_{k\in\clK} p_k \leq P_{\mathrm{sum}}^{U,\max}.
\end{equation}
The received signal at relay $m$ can be simply written as{\color{black}
\begin{equation}\label{receivedFT}
\pmb{r}_m=\sum_{\ell\in\clK} {\color{black}\sqrt{\tau p_{\ell}}} \pmb{h}_{\ell,m} s_{\ell}+\pmb{n}^{(\tau)}_{R,m},
\end{equation}}
where $\mathbf{n}^{(\tau)}_{R,m}\in{\cal CN}(0,\tau \sigma^2_R\mathbf{I}_{2N_R})$ and $\pmb{h}_{\ell,m}\in \mathbb{C}^{2N_R}$ is the vector of channels from UE $\ell$ to relay $m$.

Relay $m$ processes the received signal by applying the  beamforming matrix $\mathbf{W}_m\in \mathbb{C}^{2N_R\times 2N_R}$
for transmission:
\begin{equation}\label{Pre_FT}
\pmb{r}_{m,b}=\pmb{W}_m\pmb{r}_m=\sum_{\ell\in\clK} {\color{black}\sqrt{\tau p_{\ell}}} \pmb{W}_m\pmb{h}_{\ell,m} s_{\ell}+\pmb{W}_m \pmb{n}^{(\tau)}_{R,m}.
\end{equation}

Given the physical parameter $P^{A,\max}$ as in (\ref{IndiRelayCon})
and then $\bar{P}_R=3P^{A,\max}$, the transmit power at  relay $m$ is physically limited as
\begin{eqnarray}\label{al3}
 P_{m}^A(\pmb{p},\pmb{W}_m,\tau)={\color{black}\tau\left[\ds \sum_{\ell\in\clK} p_{\ell}||\pmb{W}_m\pmb{h}_{\ell,m}||^2+
   \sigma_R^2||\pmb{W}_m||^2\right] \leq \bar{P}_R,\
   \ m\in\clM.}
\end{eqnarray}
Given a budget  $P_{\mathrm{sum}}^{R,\max}$ as in (\ref{SumRelayCon}), the sum transmit power by the relays is also constrained as
\begin{eqnarray}\label{al4}
  P_{\mathrm{sum}}^R(\tau,\pmb{p},\pmb{W}) & =&(1-\tau)\ds \sum_{m\in\clM} P_{m}^A(\pmb{p},\pmb{W}_m,\tau) \nonumber\\
   & =&{\color{black}(1-\tau)\tau\ds\sum_{m\in\clM}\left(\sum_{\ell\in\clK} p_{\ell}||\pmb{W}_m\pmb{h}_{\ell,m}||^2+
  \sigma_R^2||\pmb{W}_m||^2\right)}\nonumber\\
  &\leq &P_{\mathrm{sum}}^{R,\max}.
\end{eqnarray}
The received signal at UE $k$ is virtually expressed as
\begin{eqnarray}\label{al5}
y_k&=&\ds\sqrt{\tau p_{a(k)}}\sum_{m\in\clM}\pmb{g}_{m,k}^T
\pmb{W}_m\pmb{h}_{a(k),m} s_{a(k)} +\sqrt{\tau p_k}\sum_{m\in\clM}\pmb{g}_{m,k}^T
\pmb{W}_m\pmb{h}_{k,m} s_{k}\nonumber\\
&&+ \ds\sum_{m\in\clM}\pmb{g}_{m,k}^T\left(\sum_{\ell\in\clK\setminus\{k,a(k)\}} \sqrt{\tau p_\ell}\pmb{W}_m\pmb{h}_{\ell,m} s_\ell+\pmb{W}_m{\color{black}\pmb{n}^{(\tau)}_{R,m}}\right)+n_k.
\end{eqnarray}

Under the definitions
\begin{equation}\label{al6}
\begin{array}{c}
{\color{black} \clL_{k,\ell}(\pmb{W})=\ds\sum_{m\in\clM}\pmb{f}_{m,k}^H \pmb{W}_m\pmb{h}_{\ell,m},} \\
\clL_k(\pmb{W})=\begin{bmatrix}\pmb{f}_{1,k}^H \pmb{W}_1&\pmb{f}_{2,k}^H \pmb{W}_2&...&\pmb{f}_{M,k}^H \pmb{W}_M
\end{bmatrix},
\end{array}
\end{equation}
the SINR at UE $k$ can be calculated as
\begin{eqnarray}
\gamma_{k}(\pmb{p}, \pmb{W},\tau)&=&{\color{black}\ds p_{a(k)}|\clL_{k,a(k)}(\pmb{W})|^2/
\left[\ds\sum_{\ell\in\clK\setminus\{k,a(k)\}} p_{\ell}|\clL_{k,\ell}(\pmb{W})|^2+\sigma_R^2||\clL_k(\pmb{W})||^2 +\sigma^2_{k}/\tau\right]}.
\label{al7}
\end{eqnarray}
Thus, the  throughput at the $k$th UE pair
is defined by the following function of beamforming matrix $\pmb{W}=\{\pmb{W}_m\}_{m\in\clM}$, power allocation vector $\pmb{p}$ and time-fraction $\tau$:
\begin{eqnarray}
R_k(\tau,\pmb{p}, \pmb{W})=(1-\tau) \ln ( 1+\gamma_{k}(\pmb{p}, \pmb{W},\tau))+
(1-\tau)\ln (1+\gamma_{a(k)}(\pmb{p}, \pmb{W},\tau)),  \ k=1,\ldots,K.\label{al8}
\end{eqnarray}
Similarly to (\ref{MaxMinPair2}), the problem of maximin exchange information  throughput optimization subject to
transmit power constraints is formulated as
\begin{subequations}\label{al10}
\begin{eqnarray}
&\ds\max_{0<\tau<1,\pmb{W}, \pmb{p}} &\ds\min_{k=1,\dots,K}
 \ds (1-\tau)\left[\ln (1+ \gamma_{k}(\pmb{p}, \pmb{W},\tau))+\ln (1+\gamma_{a(k)}(\pmb{p}, \pmb{W},\tau))\right] \label{al10a}\\
& \st & \eqref{al1}, \eqref{al2}, \eqref{al3},  \eqref{al4},\label{al10b}
\end{eqnarray}
\end{subequations}
while the problem of maximizing the network EE subject to UE QoS in terms of the exchange information throughput thresholds
is formulated similarly to (\ref{e1})  as
\begin{subequations}\label{al9}
\begin{eqnarray}
&\ds\max_{0<\tau<1,\pmb{W}, \pmb{p}} &
 \ds  {\color{black} \frac{ \ds\sum_{k=1}^K(1-\tau)\left[\ln (1+ \gamma_{k}(\pmb{p}, \pmb{W},\tau))+\ln (1+\gamma_{a(k)}(\pmb{p}, \pmb{W},\tau))\right]}{\zeta(P^U_{\rm sum}(\tau,\mathbf{p})+P^R_{\rm sum}(\tau,\mathbf{p},
\mathbf{W}))+MP^{\rm R}+2KP^{\rm U}}} \label{al9a}\\
& \st & \eqref{al1}, \eqref{al2}, \eqref{al3},  \eqref{al4}\label{al9b}\\
&&R_k(\tau,\pmb{p}, \pmb{W})\geq r_k, k=1,\cdots, K.\label{al9c}
\end{eqnarray}
\end{subequations}
The next two subsections are devoted to their computation.
\subsection{TF-wise HD TWR maximin exchange information throughput optimization}
Similarly to (\ref{MaxMinPair3}), problem (\ref{al10}) of maximin exchange information  throughput optimization
is equivalently expressed by the following optimization problem
with using new variables $\pmb{\beta}=(\beta_1,\dots, \beta_{2K})^T$ defined from (\ref{betak}):
\allowdisplaybreaks[4]
\begin{subequations}\label{al11}
\begin{eqnarray}
&\ds\max_{0<\tau<1,\pmb{W}, \atop
\pmb{\alpha}, \pmb{\beta}} &\ds\min_{k=1,\dots,K}
\left[ \ds (1-\tau)\ln (1+|\clL_{k,a(k)}(\pmb{W})|^2/\sqrt{\alpha_{k}\beta_{a(k)}})\right.\nonumber\\
 &&\hspace*{2cm}\left.+(1-\tau)\ln (1+|\clL_{a(k),k}(\pmb{W})|^2/\sqrt{\alpha_{a(k)}\beta_{k}})\right]  \label{al11a}\\
& \st & \sum_{\ell\in\clK\setminus\{k,a(k)\}} \Psi_{k,\ell}(\pmb{W}, \alpha_k, \beta_{\ell})
+{\color{black}\sigma_R^2} \Upsilon_k(\pmb{W},\alpha_k) +{\color{black}\sigma^2_{k}/\tau\sqrt{\alpha_k}} \leq 1, k\in\clK,\label{al11b} \\
&& {\color{black}\beta_k\geq 1/(\bar{P}_{\rm UE})^2}, \ k\in\clK, \label{al11c} \\
&& \sum_{k\in\clK} \tau/\sqrt{\beta_k} \leq P_{\mathrm{sum}}^{U,\max}, \label{al11d} \\
&& {\color{black}\tau\ds\left[ \sum_{\ell\in\clK} \Phi_{\ell,m}(\pmb{W}_m,1,\beta_{\ell}) +
   \sigma_R^2||\pmb{W}_m||^2\right] \leq  \bar{P}_R}, \ m\in\clM, \label{al11e} \\
&& {\color{black}(1-\tau)\tau}\ds\sum_{m\in\clM}\left(\ds \sum_{\ell\in\clK} \Phi_{\ell,m}(\pmb{W}_m,1,\beta_{\ell}) +
 {\color{black} \sigma_R^2}||\pmb{W}_m||^2\right)\leq P_{\mathrm{sum}}^{R,\max}. \label{al11f}
\end{eqnarray}
\end{subequations}
Recalling the definition (\ref{Psi}), rewrite (\ref{al11d})-(\ref{al11f}) by
\[
\begin{array}{c}
\ds\sum_{k\in\clK} 1/\sqrt{\beta_k} \leq P_{\mathrm{sum}}^{U,\max}/\tau,  \\
 \ds \sum_{\ell\in\clK} \Phi_{\ell,m}(\pmb{W}_m,1,\beta_{\ell}) +
   {\color{black}\sigma_R^2}||\pmb{W}_m||^2 \leq
 {\color{black} \bar{P}_R/\tau}, \ m\in\clM, \\
 \ds\sum_{m\in\clM}(\ds \sum_{\ell\in\clK} \Phi_{\ell,m}(\pmb{W}_m,1,\beta_{\ell}) +
  {\color{black}\sigma_R^2}||\pmb{W}_m||^2)\leq P_{\mathrm{sum}}^{R,\max}/{\color{black}(1-\tau)\tau}.
\end{array}
\]
Introduce the new variables $t_1>0$ and $t_2>0$ to express $1/\tau^2$ and $1/(1-\tau)$, which satisfy the convex constraint
\begin{equation}\label{t1}
1/\sqrt{t_1}+1/t_2\leq 1.
\end{equation}
Then, (\ref{al11}) is equivalent to
\allowdisplaybreaks[4]
\begin{subequations}\label{al11t}
\begin{eqnarray}
&\ds\max_{\pmb{W}\in\clC^{N\times N}, t_1,t_2 \atop \pmb{\alpha}\in \clR_+^{2K},
\pmb{\beta}\in \clR_+^{2K}} &\varphi(\pmb{W},\pmb{\alpha}, \pmb{\beta},t_2)\triangleq \ds\min_{k=1,\dots,K}
 \ds \left[(1/t_2)\ln \left(1+|\clL_{k,a(k)}(\pmb{W})|^2/\sqrt{\alpha_{k}\beta_{a(k)}}\right)\right.\nonumber\\
&&\hspace*{2cm}+\ds\left.(1/t_2)\ln \left(1+|\clL_{a(k),k}(\pmb{W})|^2/\sqrt{\alpha_{a(k)}\beta_{k}}\right) \right] \label{al11ta}\\
& \st & \sum_{\ell\in\clK\setminus\{k,a(k)\}} \Psi_{k,\ell}(\pmb{W}, \alpha_k, \beta_{\ell})
+ {\color{black}\sigma_R^2 \Upsilon_k(\pmb{W},\alpha_k)} +\sigma^2_{k}/{\color{black}\tau}\sqrt{\alpha_k} \leq 1,
 \label{al11tb} \\
&& \beta_k\geq {\color{black}1/(\bar{P}_{\rm UE})^2}, \hspace*{2cm}k\in\clK, \label{al11tc} \\
&& \sum_{k\in\clK} 1/\sqrt{\beta_k} \leq P_{\mathrm{sum}}^{U,\max}\sqrt{t_1}, \label{al11td} \\
&& \ds \sum_{\ell\in\clK} \Phi_{\ell,m}(\pmb{W}_m,1,\beta_{\ell}) +
  {\color{black} \sigma_R^2 ||\pmb{W}_m||^2} \leq {\color{black}\bar{P}_R\sqrt{t_1}},  m\in\clM, \label{al11te} \\
&& \ds{\color{black}\frac{1}{\sqrt{t_1}}}\sum_{m\in\clM}(\ds \sum_{\ell\in\clK} \Phi_{\ell,m}(\pmb{W}_m,1,\beta_{\ell}) +
 {\color{black}\sigma_R^2 ||\pmb{W}_m||^2})\leq t_2 P_{\mathrm{sum}}^{R,\max}, \label{al11tf}
\end{eqnarray}
\end{subequations}
where all constraints (\ref{al11tb})-(\ref{al11tf}) are convex. Therefore, the next step is to approximate the objective
function in (\ref{al11ta}).

Suppose  $(\pmb{W}^{(\kappa)}, \pmb{\alpha}^{(\kappa)}, \pmb{\beta}^{(\kappa)}, t^{(\kappa)}_1, t^{(\kappa)}_2)$ is a feasible point for (\ref{al11t}) found at the $(\kappa-1)$th iteration. Applying (\ref{ine1}) in the Appendix for
\[
x=1/|\clL_{k,a(k)}(\pmb{W})|^2, y=\sqrt{\alpha_{k}\beta_{a(k)}}, t=t_2
\]
and
\[
\bar{x}=1/|\clL_{k,a(k)}(\pmb{W}^{(\kappa)})|^2, \bar{y}=\sqrt{\alpha^{(\kappa)}_{k}\beta^{(\kappa)}_{a(k)}},
\bar{t}=t_2^{(\kappa)}
\]
and using inequality (\ref{inequa1}) yields
\begin{eqnarray}\label{al13}
(1/t_2)\ln \left(1|\clL_{k,a(k)}(\pmb{W})|^2/\sqrt{\alpha_{k}\beta_{a(k)}}\right)&\geq& \Gamma_{k,a(k)}^{(\kappa)}(\pmb{W},\alpha_{k},\beta_{a(k)}, t_2) \end{eqnarray}
over the trust region (\ref{tru1}),
for
\[
\begin{array}{lll}
x_{k,a(k)}^{(\kappa)}&=&|\clL_{k,a(k)}(\pmb{W}^{(\kappa)})|^2/\sqrt{\alpha^{(\kappa)}_{k}\beta^{(\kappa)}_{a(k)}}\\
c_{k,a(k)}^{(\kappa)}&=&(2/t_2^{(\kappa)})\ln\left(1+x_{k,a(k)}^{(\kappa)}\right) >0, \\
d_{k,a(k)}^{(\kappa)}&=&x_{k,a(k)}^{(\kappa)}/(x_{k,a(k)}^{(\kappa)}+1) t_2^{(\kappa)}>0, \\
e_{k,a(k)}^{(\kappa)}&=&(1/t_2^{(\kappa)})^2\ln\left(1+x_{k,a(k)}^{(\kappa)}\right) >0,
\end{array}
\]
and
\begin{eqnarray}
\Gamma_{k,a(k)}^{(\kappa)}(\pmb{W},\alpha_{k},\beta_{a(k)}, t_2)
&\triangleq& \nonumber\\
\ds c_{k,a(k)}^{(\kappa)}+d_{k,a(k)}^{(\kappa)}
\left[2-\frac{|\clL_{k,a(k)}(\pmb{W}^{(\kappa)})|^2}{{\color{black}2\Re\{\clL_{k,a(k)}(\pmb{W})(\clL_{k,a(k)}(\pmb{W}^{(\kappa)}))^*\}}
-|\clL_{k,a(k)}(\pmb{W}^{(\kappa)})|^2}\right.&&\nonumber\\
\ds\left. -\sqrt{\alpha_k\beta_{a(k)}}/\sqrt{\alpha^{(\kappa)}_{k}\beta^{(\kappa)}_{a(k)}}
\right]-e_{k,a(k)}^{(\kappa)} t_2.&&\label{al13a}
\end{eqnarray}
By Algorithm \ref{alg3} we propose a path-following procedure for computing (\ref{al11t}), which
solves the following convex optimization problem at the $\kappa$th iteration to generate the next feasible  point $(\pmb{W}^{(\kappa+1)}, \pmb{\alpha}^{(\kappa+1)}, \pmb{\beta}^{(\kappa+1)}, t^{(\kappa+1)}_1, t^{(\kappa+1)}_2)$:
\begin{eqnarray}\label{al15}
&\ds\max_{\pmb{W}, \pmb{\alpha}, \atop \pmb{\beta}, t_1, t_2} &
\ds \min_{k=1,\ldots ,K}
\left[G^{(\kappa)}_{k,a(k)}(\pmb{W},\alpha_{k},\beta_{a(k)}, t_1, t_2)
 +G^{(\kappa)}_{a(k),k}(\pmb{W},\alpha_{a(k)},\beta_{k}, t_1, t_2)\right] \nonumber \\
& \st & \eqref{t1}, \eqref{al11tb}, \eqref{al11tc}, \eqref{al11td}, \eqref{al11te}, \eqref{al11tf}, \eqref{tru1}.
\end{eqnarray}
Analogously to Algorithm \ref{alg1}, the sequence $\{ (\pmb{W}^{(\kappa)},\pmb{\alpha}^{(\kappa)},\pmb{\beta}^{(\kappa)},
t^{(\kappa)}_1, t^{(\kappa)}_2) \}$ generated by Algorithm \ref{alg3}
at least converges to a local optimal solution  of  (\ref{al11t}).
\begin{algorithm}
\caption{Path-following algorithm for TF-wise HD TWR exchange throughput optimization} \label{alg3}
\begin{algorithmic}
\STATE \textbf{initialization}: Set $\kappa=0$. Initialize a feasible point
$(\pmb{W}^{(0)}, \pmb{\alpha}^{(0)}, \pmb{\beta}^{(0)}, t^{(0)}_1, t^{(0)}_2))$ for the convex constraints
(\ref{al11tb})-(\ref{al11tf}) and
 $R_1=\varphi(\pmb{W}^{(0)}, \pmb{\alpha}^{(0)}, \pmb{\beta}^{(0)}, t_2^{(0)})$.
\REPEAT \STATE $\bullet$ $R_0= R_1$.
\STATE $\bullet$ Solve the
convex optimization problem \eqref{al15} to obtain the
solution $(\pmb{W}^{(\kappa+1)}, \pmb{\alpha}^{(\kappa+1)}, \pmb{\beta}^{(\kappa+1)},  t^{(\kappa+1)}_1, t^{(\kappa+1)}_2)$.
\STATE $\bullet$ Update $R_1=\varphi(\pmb{W}^{(\kappa+1)}, \pmb{\alpha}^{(\kappa+1)}, \pmb{\beta}^{(\kappa+1)},t^{(\kappa+1)}_2)$.
\STATE $\bullet$ Reset $\kappa \to\kappa+1$.
 \UNTIL{$\frac{R_1-R_0}{R_0} \leq
 \epsilon$ for given tolerance $\epsilon>0$}.
\end{algorithmic}
\end{algorithm}
\subsection{TF-wise HD TWR energy-efficiency maximization }
Similarly to (\ref{al11t}),
problem (\ref{al9}) of TF-wise HD TWR energy efficiency can be equivalently expressed by
\begin{subequations}\label{al12}
\begin{eqnarray}
\ds\max_{\pmb{W}, t_1, t_2,
 \atop  \pmb{\alpha}, \pmb{\beta}} \ \Theta(\pmb{W},\pmb{\beta},t_2 )\quad\st\quad
 \eqref{al11tb}, \eqref{al11tc}, \eqref{al11td},  \eqref{al11te},  \eqref{al11tf},  \hspace*{3cm}\label{al12a}\\
\ds \ln \left(1+|\clL_{k,a(k)}(\pmb{W})|^2/\sqrt{\alpha_{k}\beta_{a(k)}}\right)
+\ln \left(1+ |\clL_{a(k),k}(\pmb{W})|^2/\sqrt{\alpha_{a(k)}\beta_{k}}\right)
 \geq t_2 r_k,\ \label{al12c}\\
 k=1,\cdots, K,\nonumber
\end{eqnarray}
\end{subequations}
where
\[
\Theta(\pmb{W},\pmb{\beta},t_2 )\triangleq
 \ds \ds\sum_{k=1}^K\left[\ln (1+\frac{|\clL_{k,a(k)}(\pmb{W})|^2}{\sqrt{\alpha_{k}\beta_{a(k)}}} )
 +\ln (1+ \frac{|\clL_{a(k),k}(\pmb{W})|^2}{\sqrt{\alpha_{a(k)}\beta_{k}}} )\right]/t_2\pi(\pmb{\beta},\pmb{W},
 t_1)
\]
with the consumption power function $\pi(\pmb{\beta},\pmb{W})$ defined by
\begin{eqnarray}\label{al16}
\pi(\pmb{\beta},\pmb{W}, t_1) &\triangleq& \zeta\left[\sum_{k\in\clK}1/\sqrt{\beta_k t_1} + (1-1/\sqrt{t_1})\ds\sum_{m\in\clM}(\ds \sum_{\ell\in\clK} \Phi_{\ell,m}(\pmb{W}_m,1,\beta_{\ell}) +
  \sigma_R^2||\pmb{W}_m||^2/\sqrt{t_1})\right]\nonumber\\
  && + MP^{\rm R}+2KP^{\rm U}.
\end{eqnarray}
Using the inequalities
\begin{eqnarray*}
\frac{\Phi_{\ell,m}(\pmb{W}_m,1,\beta_{\ell})}{\sqrt{t_1}} &\geq& \frac{\Phi_{\ell,m}(\pmb{W}_m^{(\kappa)},1,\beta_{\ell}^{(\kappa)})}{\sqrt{t_1^{(\kappa)}}}
+ 2 \langle \frac{(\pmb{W}_m^{(\kappa)})^H \pmb{h}_{\ell,m} \pmb{h}_{\ell,m}^H }{\sqrt{\beta_{\ell}^{(\kappa)} t_1^{(\kappa)} }},\pmb{W}_m-\pmb{W}_m^{(\kappa)}  \rangle \nonumber \\
&& - \frac{\Phi_{\ell,m}(\pmb{W}_m^{(\kappa)},1,\beta_{\ell}^{(\kappa)})}{2 (t_1^{(\kappa)})^{3/2}}(t_1 - t_1^{(\kappa)} )
- \frac{|| \pmb{h}_{\ell,m}^H \pmb{W}_m^{(\kappa)} ||^2 }{2 \sqrt{t_1^{(\kappa)}} (\beta_{\ell}^{(\kappa)})^{3/2}}(\beta_{\ell} - \beta_{\ell}^{(\kappa)})
\end{eqnarray*}
and
\[
\frac{||\pmb{W}_m||^2}{t_1} \geq \frac{||\pmb{W}_m^{(\kappa)}||^2}{t_1^{(\kappa)}} + 2 \langle \frac{\pmb{W}_m^{(\kappa)}}{t_1^{(\kappa)}}, \pmb{W}_m -  \pmb{W}_m^{(\kappa)} \rangle - \frac{||\pmb{W}_m^{(\kappa)}||^2}{(t_1^{(\kappa)})^2} (t_1 - t_1^{(\kappa)})
\]
which follow from the convexity of functions defined in (\ref{Psi}), one can obtain
\begin{equation}\label{al16T}
\pi(\pmb{\beta},\pmb{W}, t_1) \leq \pi^{(\kappa)}(\pmb{\beta},\pmb{W}, t_1),
\end{equation}
where
\begin{eqnarray*}
\pi^{(\kappa)}(\pmb{\beta},\pmb{W}, t_1)&\triangleq&\zeta\ds\left[ \sum_{k\in\clK} \frac{1}{\sqrt{\beta_k t_1}} + \ds\sum_{m\in\clM}(\ds \sum_{\ell\in\clK} \Phi_{\ell,m}(\pmb{W}_m,1,\beta_{\ell}) +  \sigma_R^2\frac{||\pmb{W}_m||^2}{\sqrt{t_1}}) \right.\nonumber \\
&& - \sum_{m\in\clM} \sum_{\ell\in\clK}(\frac{\Phi_{\ell,m}(\pmb{W}_m^{(\kappa)},1,\beta_{\ell}^{(\kappa)})}{\sqrt{t_1^{(\kappa)}}}
 + 2 \langle \frac{(\pmb{W}_m^{(\kappa)})^H \pmb{h}_{\ell,m} \pmb{h}_{\ell,m}^H }{\sqrt{\beta_{\ell}^{(\kappa)} t_1^{(\kappa)} }},\pmb{W}_m-\pmb{W}_m^{(\kappa)}  \rangle \nonumber \\
&& - \frac{\Phi_{\ell,m}(\pmb{W}_m^{(\kappa)},1,\beta_{\ell}^{(\kappa)})}{2 (t_1^{(\kappa)})^{3/2}}(t_1 - t_1^{(\kappa)} )
 - \frac{|| \pmb{h}_{\ell,m}^H \pmb{W}_m^{(\kappa)} ||^2 }{2 \sqrt{t_1^{(\kappa)}} (\beta_{\ell}^{(\kappa)})^{3/2}}(\beta_{\ell} - \beta_{\ell}^{(\kappa)}) \nonumber \\
&&\ds\left. - \sum_{m\in\clM} \sigma_R^2 (\frac{||\pmb{W}_m^{(\kappa)}||^2}{t_1^{(\kappa)}} + 2 \langle \frac{\pmb{W}_m^{(\kappa)}}{t_1^{(\kappa)}}, \pmb{W}_m -  \pmb{W}_m^{(\kappa)} \rangle - \frac{||\pmb{W}_m^{(\kappa)}||^2}{(t_1^{(\kappa)})^2} (t_1 - t_1^{(\kappa)}))\right] \nonumber \\
&& + MP^{\rm R}+2KP^{\rm U}, \nonumber 
\end{eqnarray*}
which is a convex function.

Suppose that $(\pmb{W}^{(\kappa)},\pmb{\alpha}^{(\kappa)},\pmb{\beta}^{(\kappa)}, t_1^{(\kappa)}, t_2^{(\kappa)})$ is a feasible point for (\ref{al12}) found from the ($\kappa-1$)th iteration.
Applying inequality (\ref{ine2}) in the Appendix for
\[
x=1/|\clL_{k,a(k)}(\pmb{W})|^2, y=\sqrt{\alpha_{k}\beta_{a(k)}}, z=\pi(\pmb{\beta},\pmb{W}, t_1),
t=t_2
\]
and
\[
\bar{x}=1/|\clL_{k,a(k)}(\pmb{W}^{(\kappa)})|^2, \bar{y}=\sqrt{\alpha^{(\kappa)}_{k}\beta^{(\kappa)}_{a(k)}},
\bar{z}=\pi(\pmb{\beta}^{(\kappa)},\pmb{W}^{(\kappa)}, t_1^{(\kappa)}), \bar{t}=t_2^{(\kappa)}
\]
and using inequality (\ref{inequa1})
yield
\begin{eqnarray}\label{al17}
\ds\frac{\ln \left(1+|\clL_{k,a(k)}(\pmb{W})|^2/\sqrt{\alpha_{k}\beta_{a(k)}}\right)}{\color{black}t_2\pi(\pmb{\beta},\pmb{W}, t_1) }&\geq&
\tilde{F}^{(\kappa)}_{k,a(k)}(\pmb{W}, \alpha_{k}, \pmb{\beta}, t_2)
\end{eqnarray}
over the trust region (\ref{tru1})
for
\begin{equation}\label{al18}
\begin{array}{lll}
x_{k,a(k)}^{(\kappa)}&=&|\clL_{k,a(k)}(\pmb{W}^{(\kappa)})|^2/\sqrt{\alpha^{(\kappa)}_{k}\beta^{(\kappa)}_{a(k)}},\\
p_{k,a(k)}^{(\kappa)}&=&3\ds\left[\ln(1+x_{k,a(k)}^{(\kappa)})\right]/t_2^{(\kappa)} t^{(\kappa)} >0, \\
q_{k,a(k)}^{(\kappa)}&=&x_{k,a(k)}^{(\kappa)}/(x_{k,a(k)}^{(\kappa)}+1) t_2^{(\kappa)} t^{(\kappa)}>0, \\
r_{k,a(k)}^{(\kappa)}&=&\ds \left[\ln(1+x_{k,a(k)}^{(\kappa)})\right]/(t_2^{(\kappa)})^2 t^{(\kappa)} >0, \\
s_{k,a(k)}^{(\kappa)}&=&\ds \left[\ln(1+x_{k,a(k)}^{(\kappa)})\right]/t_2^{(\kappa)} (t^{(\kappa)})^2 >0,
\end{array}
\end{equation}
and
\begin{eqnarray}
\tilde{F}^{(\kappa)}_{k,a(k)}(\pmb{W}, \alpha_{k}, \pmb{\beta}, t_2 )&\triangleq&\nonumber\\
\ds p^{(\kappa)}_{k,a(k)} + q^{(\kappa)}_{k,a(k)}
\left[2-\frac{|\clL_{k,a(k)}(\pmb{W}^{(\kappa)})|^2}{{\color{black}2\Re\{\clL_{k,a(k)}(\pmb{W})(\clL_{k,a(k)}(\pmb{W}^{(\kappa)}))^*\}}
-|\clL_{k,a(k)}(\pmb{W}^{(\kappa)})|^2}\right.&&\nonumber\\
\ds\left.-\sqrt{\alpha_k\beta_{a(k)}}/\sqrt{\alpha^{(\kappa)}_{k}\beta^{(\kappa)}_{a(k)}}\right]
- r^{(\kappa)}_{k,a(k)} t_2 - s^{(\kappa)}_{k,a(k)}{\pi^{(\kappa)}(\pmb{\beta},\pmb{W}, t_1)}.
\label{al17a}
\end{eqnarray}
By Algorithm \ref{alg4} we propose a path-following procedure for computing (\ref{al12}), which
solves the following convex optimization problem at the $\kappa$th iteration to generate the next feasible  point
$(\pmb{W}^{(\kappa+1)},\pmb{\alpha}^{(\kappa+1)},\pmb{\beta}^{(\kappa+1)}, t_1^{(\kappa+1)}, t_2^{(\kappa+1)})$:
\begin{subequations}\label{ConvexOpt.FTee}
\begin{eqnarray}
& \ds\max_{\pmb{W}, t_1, t_2  \atop \pmb{\alpha}, \pmb{\beta} } &
\ds \sum_{k=1}^K
[\tilde{F}^{(\kappa)}_{k,a(k)}(\pmb{W},\alpha_{k},\pmb{\beta}, t_2)
+\tilde{F}^{(\kappa)}_{a(k),k}(\pmb{W},\alpha_{a(k)},\pmb{\beta}, t_2)]\label{FTee.a}\\
& \st & \eqref{t1}, (\ref{al11tb})-(\ref{al11tf} ), (\ref{tru1}),\label{FTee.b}\\
&&f^{(\kappa)}_{k,a(k)}(\pmb{W},\alpha_{k},\beta_{a(k)}) +f^{(\kappa)}_{a(k),k}(\pmb{W},\alpha_{a(k)},\beta_{k})
\geq t_2 r_k, k=1,\cdots, K,\label{FTee.c}
\end{eqnarray}
\end{subequations}
where $f^{(\kappa)}_{k,a(k)}$ are defined from (\ref{ap1}).

Analogously to Algorithm \ref{alg1}, the sequence $\{ (\pmb{W}^{(\kappa)},\pmb{\alpha}^{(\kappa)},\pmb{\beta}^{(\kappa)},
t^{(\kappa)}_1, t^{(\kappa)}_2) \}$ generated by Algorithm \ref{alg4}
at least converges to a local optimal solution  of  (\ref{al12}).
\begin{algorithm}
\caption{Path-following algorithm for TF-wise HD TWR energy-efficiency optimization} \label{alg4}
\begin{algorithmic}
\STATE \textbf{initialization}: Set $\kappa=0$. Initialize a feasible point
$(\pmb{W}^{(0)}, \pmb{\alpha}^{(0)}, \pmb{\beta}^{(0)}, t^{(0)}_1, t^{(0)}_2))$ for the convex constraints
(\ref{al12a})-(\ref{al12c}) and
 $e_1=\Theta(\pmb{W}^{(0)}, \pmb{\alpha}^{(0)}, \pmb{\beta}^{(0)}, t_2^{(0)})$.
\REPEAT \STATE $\bullet$ $e_0= e_1$.
\STATE $\bullet$ Solve the
convex optimization problem \eqref{ConvexOpt.FTee} to obtain the
solution $(\pmb{W}^{(\kappa+1)}, \pmb{\alpha}^{(\kappa+1)}, \pmb{\beta}^{(\kappa+1)},  t^{(\kappa+1)}_1, t^{(\kappa+1)}_2)$.
\STATE $\bullet$ Update $e_1=\Theta(\pmb{W}^{(\kappa+1)}, \pmb{\alpha}^{(\kappa+1)}, \pmb{\beta}^{(\kappa+1)},t^{(\kappa+1)}_2)$.
\STATE $\bullet$ Reset $\kappa \to\kappa+1$.
 \UNTIL{$\frac{e_1-e_0}{e_0} \leq
 \epsilon$ for given tolerance $\epsilon>0$}.
\end{algorithmic}
\end{algorithm}

An initial feasible point $(\pmb{W}^{(0)}, \pmb{\alpha}^{(0)}, \pmb{\beta}^{(0)}, t_1^{(0)}, t_2^{(0)})$ for initializing
Algorithm \ref{alg4} can be found by using Algorithm \ref{alg3} for
computing (\ref{al11t}), which terminates upon
\begin{equation}\label{inipoint4}
\ds{\min_{k=1,\ldots ,K}} \left[\ln \left(1+|\clL_{k,a(k)}(\pmb{W})|^2/\sqrt{\alpha_{k}\beta_{a(k)}}\right)
+\ln \left(1+|\clL_{a(k),k}(\pmb{W})|^2/\sqrt{\alpha_{a(k)}\beta_{k}}\right) \right]/
t_2r_k\geq 1
\end{equation}
to satisfy (\ref{al12a})-(\ref{al12c}).

\section{Numerical Results}\label{sec:Simulation}
In this section, simulation results are presented to demonstrate the advantage of the TF-wise HD TWR conisered
in Section III  over FD-based TWR considered in Section II and HD TWR considered in \cite{STDP17}.
The channel $\mathbf{h}_{k,m}$ from UE $\ell$ to relay $m$ and
the channel $\mathbf{g}_{m,k}$ from relay $m$ to UE $k$  are assumed  Rayleigh fading, which
are modelled by independent circularly-symmetric complex Gaussian random variables
with zero means and unit variances.
The power of the background noises $\pmb{n}_{R,m}$ at relay $m$
and $n_k$ at UE $k$ are normalized to $\sigma_R^2=\sigma_k^2=1$.
The tolerance for the  algorithms \ref{alg1}-\ref{alg4} is set as $\epsilon=10^{-4}$.
Each point of the numerical results is the average of $1,000$ random channel realizations.
Other settings are: $P_{\mathrm{sum}}^{U,\max}=KP^{U, \max}$ and $P_{\mathrm{sum}}^{R,\max}=MP^{A, \max}/2$, where
$P^{U, \max}$ and $P_{\mathrm{sum}}^{R,\max}$ are fixed at $10$ dBW and $15$ dBW; the drain efficiency
of power amplifier $1/\zeta$ is $40 \%$; the circuit powers
of each antenna in relay and UE are $0.97$ dBW and $-13$ dBW.

The scenarios of $K \in \{2,3\}$ pairs and $(M,N_R) \in \{(1, 8), (2, 4), (4, 2)\}$ are simulated.
\subsection{Maximin  exchange information throughput optimization}
\begin{figure}
\centering
\includegraphics[width=4.5in]{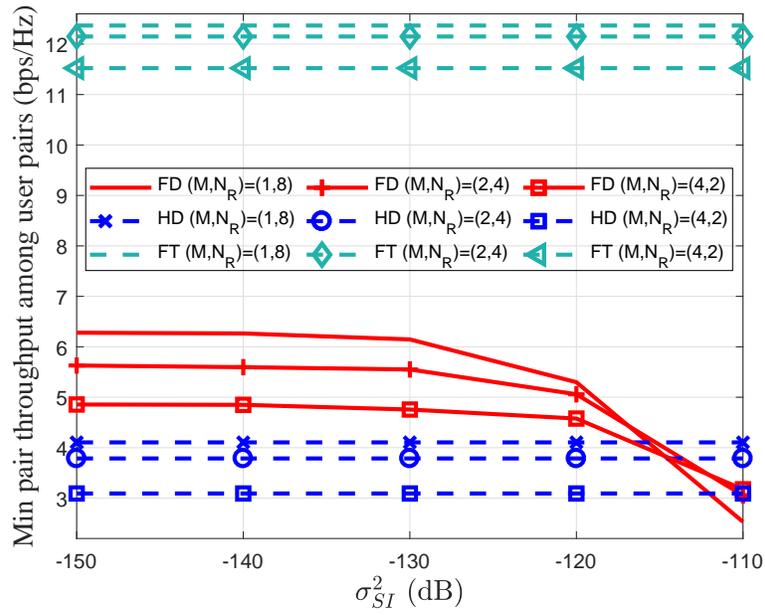}
\caption{Minimum pair exchange throughput  versus $\sigma_{SI}^2$ with $K=2$.}
\label{fig:Ex1_rate_K2}
\end{figure}

\begin{figure}
\centering
\includegraphics[width=4.5in]{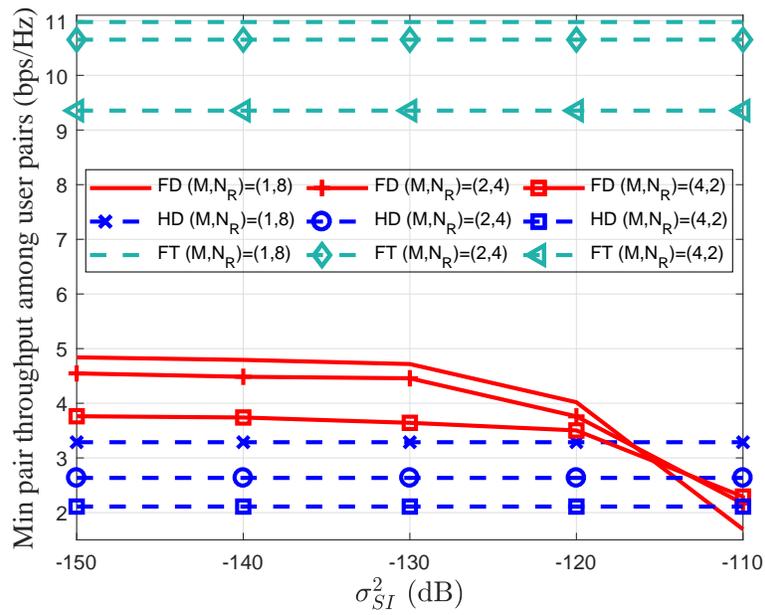}
\caption{Minimum pair exchange  throughput  versus $\sigma_{SI}^2$ with $K=3$.}
\label{fig:Ex1_rate_K3}
\end{figure}

To confirm the negative effect of the FD SI attenuation level $\sigma_{SI}$,
Fig. \ref{fig:Ex1_rate_K2} and \ref{fig:Ex1_rate_K3} plot the achievable minimum pair exchange
throughput versus SI $\sigma_{SI}^2$ with $K \in \{2,3\}$.
For small $\sigma_{SI}$ that make FD SI to the level of the background noise, the minimum pair
exchange throughput achieved by FD-based TWR
still enjoys the gain offered by FD as   is better than that obtained by HD TWR.
However, FD cannot offset for larger $\sigma_{SI}$ that make FD SI larger than the background noise,
so the former becomes worse than the latter.
In contrast, the minimum pair exchange throughput by TF-wise HD TWR is free of FD SI and it is
significantly  better than  that achieved by the other two. Certainly, using all antennas for
separated reception and transmission in time fractions within the time unit is not only much easier
implemented but is much better than FD with simultaneous reception and transmission. It has been also
shown in \cite{Naetal17a} and \cite{Naetal17b} that separated information and energy transfer in time fractions
within the unit time is much more efficient and secured than the simultaneous information and energy transfer.

 Table \ref{tab:Ex1_Ite_FDK2} and \ref{tab:Ex1_Ite_FDK3} provide a computational experience in implementing
Algorithm \ref{alg1}, which converges in less than $23$ and $36$ iterations in all considered FD SI scenarios for solving (\ref{MaxMinPair2}) with $K=2$ and $K=3$, respectively. A computational experience in implementing Algorithm \ref{alg3} is provided
by Table \ref{tab:Ex1_Ite_FT}, which shows that
Algorithm \ref{alg3} converges in less than $25$ iterations for solving (\ref{al10}) with $K=2$ and $K=3$.

\begin{table*}[!t]
   \centering
   \caption{Average number of iterations for computing (\ref{MaxMinPair2}) by Algorithm \ref{alg1} with $K=2$.}
   \begin{tabular}{ | c | c | c | c | c | c | }
    \hline
   $\sigma_{SI}^2$ (dB) & -150  &  -140 &  -130 &  -120 &  -110  \\
   \hline
   $(K, M,N_R)=(2,1,8)$ & 13.02 & 14.36 & 13.24 & 15.69 & 18.83 \\
    \hline
   $(K, M,N_R)=(2,2,4)$ & 18.16 & 17.92 & 16.80 & 17.06 & 19.53  \\
  \hline
  $(K, M,N_R)=(2,4,2)$ & 23.25 & 18.03 & 21.09 & 19.57 & 21.61 \\
  \hline
\end{tabular}
\label{tab:Ex1_Ite_FDK2}
\end{table*}

\begin{table*}[!t]
   \centering
   \caption{Average number of iterations for computing (\ref{MaxMinPair2}) by Algorithm \ref{alg1} with $K=3$.}
   \begin{tabular}{ | c | c | c | c | c | c | }
    \hline
   $\sigma_{SI}^2$ (dB) & -150  &  -140 &  -130 &  -120 &  -110  \\
   \hline
   $(K, M,N_R)=(3,1,8)$ & 30.49 & 27.81 & 30.26 & 35.76 & 26.22  \\
    \hline
   $(K, M,N_R)=(3,2,4)$ & 24.86 & 26.02 & 26.31 & 27.05 & 31.33  \\
  \hline
  $(K, M,N_R)=(3,4,2)$ & 36.10 & 24.85 & 33.47 & 34.35 & 22.96 \\
  \hline
\end{tabular}
\label{tab:Ex1_Ite_FDK3}
\end{table*}

\begin{table*}[!t]
   \centering
   \caption{Average number of iterations for computing (\ref{al10}) by Algorithm \ref{alg3}.}
   \begin{tabular}{ | c | c | c |  }
    \hline
    Iterations & K=2 & K=3 \\
    \hline
    $(M,N_R)=(1,8)$ & 23.55 & 22.42 \\
    \hline
    $(M,N_R)=(2,4)$ & 25.64 & 25.75 \\
    \hline
    $(M,N_R)=(4,2)$ & 25.32 & 21.43 \\
    \hline
\end{tabular}
\label{tab:Ex1_Ite_FT}
\end{table*}

\subsection{EE maximization}
To include a comparison with HD TWR \cite{STDP17},
the exchange throughput threshold $r_k$ in (\ref{e1}) and (\ref{al9})
is set as the half of the optimal value of the maximin exchange throughput optimization problem for
HD TWR that is computed by \cite[Alg. 1]{STDP17}.

\begin{figure}
\centering
\includegraphics[width=4.5in]{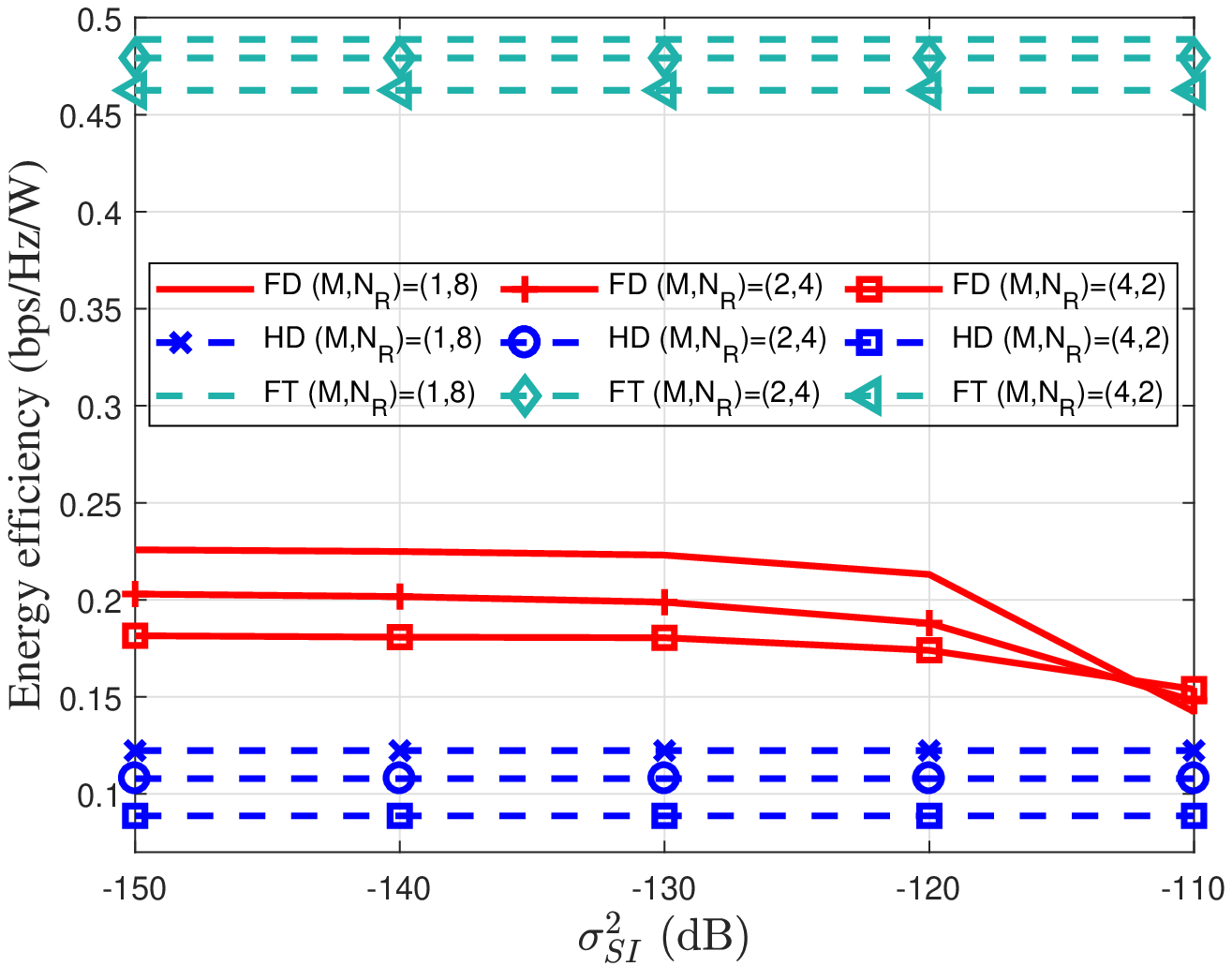}
\caption{Energy efficiency versus $\sigma_{SI}^2$ with $K=2$.}
\label{fig:Ex2_ee_K2}
\end{figure}

\begin{figure}
\centering
\includegraphics[width=4.5in]{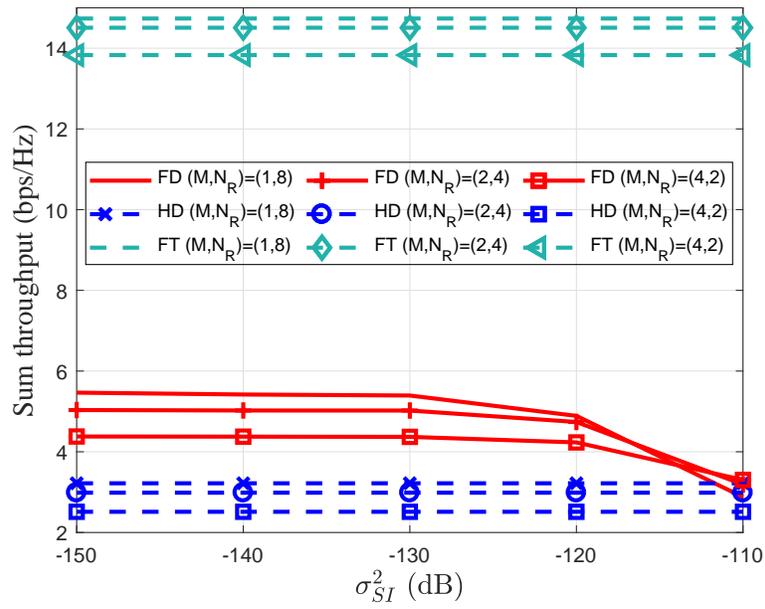}
\caption{Sum thoughput versus $\sigma_{SI}^2$ with $K=2$.}
\label{fig:Ex2_rate_K2}
\end{figure}

\begin{figure}
\centering
\includegraphics[width=4.5in]{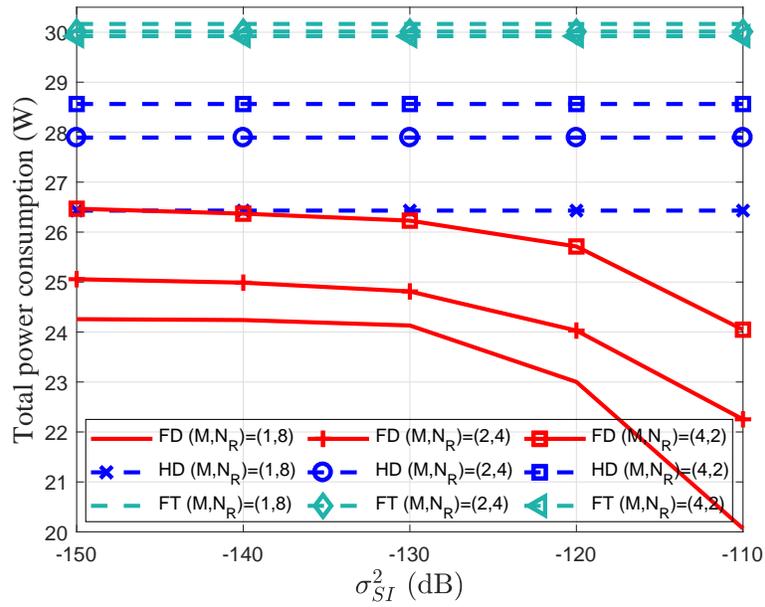}
\caption{Total power versus $\sigma_{SI}^2$ with $K=2$.}
\label{fig:Ex2_power_K2}
\end{figure}

\begin{figure}
\centering
\includegraphics[width=4.5in]{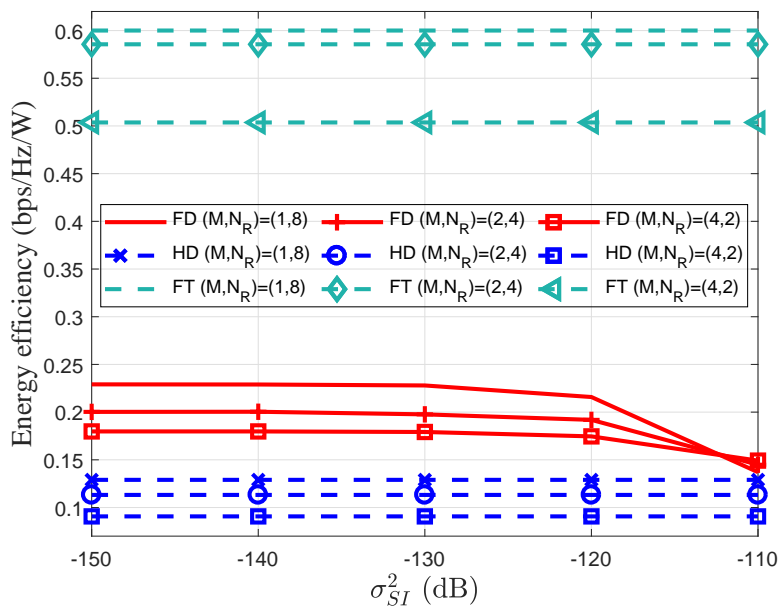}
\caption{Energy efficiency versus $\sigma_{SI}^2$ with $K=3$.}
\label{fig:Ex2_ee_K3}
\end{figure}

Fig. \ref{fig:Ex2_ee_K2} plots the energy efficiency by the three schemes
for $K=2$. As expected,  the two other schemes cannot compete with FT-wise HD TWR.
The corresponding sum throughput and transmit power plotted in Figs. \ref{fig:Ex2_rate_K2} and \ref{fig:Ex2_power_K2}
particularly explain  the superior performance of TF-wise HD TWR.
The sum throughput achieved by TF-wise HD TWR is  more than double that
achieved by FD-based TWR and HD TWR thanks to its using more power for the relay beamforming. In contrast,
Fig. \ref{fig:Ex2_power_K2} shows that
the transmit power in FD-based TWR must be controlled to make sure that its transmission does not
so severely interfere its reception. Nevertheless, FD-based TWR always achieves
better EE than  HD TWR in the considered range of $\sigma_{SI}^2$ though the gap becomes narrower as $\sigma_{SI}^2$.
For small $\sigma_{SI}^2$, FD-based TWR achieves higher sum throughput with less transmit power as compared to
HD TWR.  For larger $\sigma_{SI}^2$, the former achieves almost the same sum through as the latter does but
with much less transmission  power, keeping its EE higher than the latter.
Fig. \ref{fig:Ex2_ee_K3} for $K=3$ follows a similar pattern.

Lastly, Table \ref{tab:Ex2_Ite_FDK2}, \ref{tab:Ex2_Ite_FDK3} and \ref{tab:Ex2_Ite_FT} provide a computational experience in
implementing Algorithm \ref{alg2} for solving (\ref{e1}) and Algorithm \ref{alg4} for solving (\ref{al9}).
Algorithm \ref{alg2} needs less than $29$ and $40$ iterations on average for $K=2$ and $K=3$,
while Algorithm \ref{alg4} need less than $23$ and $24$ iterations.

\begin{table*}[!t]
   \centering
   \caption{Average number of iterations for computing (\ref{e1}) by Algorithm \ref{alg3} with $K=2$.}
   \begin{tabular}{ | c | c | c | c | c | c | }
    \hline
   $\sigma_{SI}^2$ (dB) & -150  &  -140 &  -130 &  -120 &  -110  \\
   \hline
   $(K, M,N_R)=(2,1,8)$ & 24.85 & 26.18 & 21.02 & 26.63 & 29.43 \\
    \hline
   $(K, M,N_R)=(2,2,4)$ & 26.49 & 27.76 & 26.04 & 24.18 & 27.09  \\
  \hline
  $(K, M,N_R)=(2,4,2)$ & 23.87 & 23.24 & 24.31 & 24.65 & 22.83 \\
  \hline
\end{tabular}
\label{tab:Ex2_Ite_FDK2}
\end{table*}

\begin{table*}[!t]
   \centering
   \caption{Average number of iterations for solving (\ref{e1}) by Algorithm \ref{alg3} with $K=3$.}
   \begin{tabular}{ | c | c | c | c | c | c | }
    \hline
   $\sigma_{SI}^2$ (dB) & -150  &  -140 &  -130 &  -120 &  -110  \\
   \hline
   $(K, M,N_R)=(3,1,8)$ & 29.40 & 28.59 & 30.42 & 37.31 & 40.46 \\
    \hline
   $(K, M,N_R)=(3,2,4)$ & 27.81 & 28.17 & 30.65 & 32.45 & 31.19  \\
  \hline
  $(K, M,N_R)=(3,4,2)$ &  31.75 & 24.44 & 26.13 & 25.37 & 30.38 \\
  \hline
\end{tabular}
\label{tab:Ex2_Ite_FDK3}
\end{table*}

\begin{table*}[!t]
   \centering
   \caption{Average number of iterations for computing (\ref{al9}) by Algorithm \ref{alg4}.}
   \begin{tabular}{ | c | c | c |  }
    \hline
    Iterations & K=2 & K=3 \\
    \hline
    $(M,N_R)=(1,8)$ &  20.25 & 19.38 \\
    \hline
    $(M,N_R)=(1,8)$ &  21.51 & 21.19 \\
    \hline
    $(M,N_R)=(1,8)$ &  23.13 & 24.08 \\
    \hline
\end{tabular}
\label{tab:Ex2_Ite_FT}
\end{table*}

\section{Conclusions}\label{sec:Conclusion}
This paper has considered two possible approaches for multiple pairs of users to exchange information via multiple
relays within one time slot. The first approach is based on full-duplexing (FD) at the users and relays, while the second
approach is based on  separated time-fraction-wise (TF-wise) half-duplexing (HD)
signal transmission and reception by the users and relays. It is  much easier to implement the second approach than the first approach. In order to compare their capability, we have considered two
fundamental problems of joint design of  UE power allocation and relay beamforming
to optimize the spectral efficiency and energy efficiency. Path-following optimization algorithms have
been devised  for their computation. Simulation results have confirmed their rapid convergence. TF-wise HD TWR has been shown
to easily outperform FD-based TWR and HD TWR in all aspects.
\section*{Appendix}
Let $\mbR^N_+\triangleq \{(x_1,\dots,x_N) :\ x_i>0,\ i=1,2,\ldots,N\}$ and $\mbR_+\triangleq (0,+\infty)$.
In \cite{Shetal17o}, it was proved that function $\psi(x,y,t)=(\ln (1+1/xy))/t$ is  convex on
$\mbR^3_+$. Therefore \cite{Tuybook}
\begin{eqnarray}
\ds \frac{\ln(1+1/xy)}{t}&=&\psi(x,y,t)\nonumber\\
&\geq& \psi(\bar{x},\bar{y},\bar{t})+\la \nabla \psi(\bar{x},\bar{y},\bar{t}), (x,y,t)-(\bar{x},\bar{y},\bar{t})\ra \nonumber\\
&=&2\frac{\ln(1+1/\bar{x}\bar{y})}{\bar{t}}+\frac{1}{(\bar{x}\bar{y}+1)\bar{t}}(2-
\frac{x}{\bar{x}}-\frac{y}{\bar{y}})-\frac{\ln(1+1/\bar{x}\bar{y})}{\bar{t}^2}t\label{ine1}\\
&&\quad\forall\ (x,y,t)\in\mbR^3_+, (\bar{x},\bar{y},\bar{t})\in\mbR^3_+.\nonumber
\end{eqnarray}
The right-hand-side (RHS) of (\ref{ine1})  agrees with the left-hand-side (LHS) at $(\bar{x},\bar{y},\bar{t})$.\\
Particularly,
\begin{eqnarray}
\ln(1+1/xy)\geq2\ln(1+1/\bar{x}\bar{y})+\frac{1}{(\bar{x}\bar{y}+1)}(2-
\frac{x}{\bar{x}}-\frac{y}{\bar{y}})\nonumber\\
\forall\ (x, y)\in\mbR^2_+, (\bar{x}, \bar{y})\in\mbR^2_+.\label{ine1p}
\end{eqnarray}
\begin{mylem}\label{convexlem} If function $f(\bx,t)$ is convex in $\bx$ and $t\in\mbR_+$ and also is decreased in $t$ then
function $f(\bx, \sqrt{yz})$ is convex in $\bx$ and $(y,z)\in\mbR^2_+$.
\end{mylem}
\Prf Since $\sqrt{yz}$ is a concave function, it is true that
\[
\begin{array}{r}
\sqrt{(\alpha y_1+\beta y_2)(\alpha_1 z_1+\alpha_2 z_2)}\geq \alpha_1\sqrt{y_1z_1}+\alpha_2\sqrt{y_2z_2}\\
\forall\ \alpha_i\geq 0, \alpha_1+\alpha_2=1, y_i\geq 0, z_i\geq 0, i=1, 2.
\end{array}
\]
Therefore
\[
\begin{array}{lll}
f(\alpha_1\bx_1+\alpha_2\bx_2, \sqrt{(\alpha_1y_1+\alpha_2y_2)(\alpha_1z_1+\alpha_2z_2})&\leq&
f(\alpha_1\bx_1+\alpha_2\bx_2, \alpha_1\sqrt{y_1z_1}+\alpha_2\sqrt{y_2z_2})\\
&\leq& \alpha_1f(\bx_1,\sqrt{y_1z_1})+\alpha_2f(\bx_2,\sqrt{y_2z_2}),
\end{array}
\]
showing the convexity of $f(\bx,\sqrt{yz})$.

\begin{mylem}\label{convexlem1} Function $f(x,y,t)=(\ln (1+1/xy))/t^2$ is
convex on $\mbR^3_+$.
\end{mylem}
\Prf  One has
\begin{eqnarray}
\nabla^2 f(x,y,t)&=&
\begin{bmatrix}
\ds\frac{2xy+1}{x^2(xy+1)^2t^2}&\ds\frac{1}{(xy+1)^2t^2}&\ds\frac{2}{t^3(xy+1)x}\cr
\ds\frac{1}{(xy+1)^2t^2}&\ds\frac{2xy+1}{y^2(xy+1)^2t^2}&\ds\frac{2}{t^3(xy+1)y}\cr
\ds\frac{2}{t^3(xy+1)x}&\ds\frac{2}{t^3(xy+1)y}&\ds\frac{6\ln(1+1/xy)}{t^4}
\end{bmatrix}\nonumber\\
&\succeq&
(x^2y^2(xy+1)^2t^4)^{-1}\nonumber\\
&&\times\begin{bmatrix}
 (2xy+1)y^2t^2&x^2y^2t^2&2t(xy+1)xy^2\cr
x^2y^2t^2&(2xy+1)x^2t^2&2t(xy+1)x^2y\cr
2t(xy+1)xy^2&2t(xy+1)x^2y&6(xy+1)x^2y^2
\end{bmatrix},\label{mat1}
\end{eqnarray}
because $\ln(1+1/t)\geq 1/(t+1) \quad\forall\ t>0$ \cite[Lemma 1]{Shetal17o}. Here and after $\bA\succeq \bB$ for
real symmetric matrices $\bA$ and $\bB$ means that $\bA-\bB$ is positive definite.

Then, calculating the subdeterminants of the matrix in the RHS of (\ref{mat1}) yields
\[
\begin{array}{c}
(2xy+1)y^2t^2>0,\\
\left|\begin{array}{cc}
(2xy+1)y^2t^2&x^2y^2t^2\cr
x^2y^2t^2&(2xy+1)x^2t^2
\end{array}\right|=x^2y^2t^4(3x^2y^2+4xy+1)>0,
\end{array}
\]
and
\[
\left|\begin{array}{ccc}
 (2xy+1)y^2t^2&x^2y^2t^2&2t(xy+1)xy^2\cr
x^2y^2t^2&(2xy+1)x^2t^2&2t(xy+1)x^2y\cr
2t(xy+1)xy^2&2t(xy+1)x^2y&6(xy+1)x^2y^2
\end{array}\right|=12(xy+1)^2x^5y^5t^4>0.
\]
Therefore the matrix in the RHS of (\ref{mat1}) is positive definite, implying that the Hessian $\nabla^2 f(x,y,t)$
is positive definite too, which is the necessary and sufficient condition for the convexity of $f$ \cite{Tuybook}.\qed
By applying Lemma \ref{convexlem1} and Lemma \ref{convexlem}, function $\psi(x,y,z,t)=(\ln (1+1/xy))/zt$ is  convex on
 $\mbR^4_+$. Therefore \cite{Tuybook}
\begin{eqnarray}
\ds \frac{\ln(1+1/xy)}{zt}&=&\psi(x,y,z,t)\nonumber\\
&\geq& \psi(\bar{x},\bar{y},\bar{z},\bar{t})+\la \nabla \psi(\bar{x},\bar{y},\bar{z},\bar{t}), (x,y,z,t)-(\bar{x},\bar{y},\bar{z},\bar{t})\ra\nonumber\\
&=&\ds
3\frac{\ln(1+1/\bar{x}\bar{y})}{\bar{z}\bar{t}}+
\frac{1}{(\bar{x}\bar{y}+1)\bar{z}\bar{t}}(2-\frac{x}{\bar{x}}-\frac{y}{\bar{y}})\nonumber\\
&&\ds -\frac{\ln(1+1/\bar{x}\bar{y})}{\bar{z}^2\bar{t}}z-\frac{\ln(1+1/\bar{x}\bar{y})}{\bar{z}\bar{t}^2}t\label{ine2}\\
&&\quad\forall\ (x, y, z, t)\in\mbR^4_+, (\bar{x}, \bar{y}, \bar{z}, \bar{t})\in\mbR^4_+.\nonumber
\end{eqnarray}
\bibliographystyle{ieeetr}
\bibliography{TwoWayRelay_refs}
\end{document}